\documentclass[notitlepage]{revtex4-2}

\usepackage[utf8]{inputenc}
\usepackage[dvipsnames]{xcolor}
\usepackage[colorlinks=true,citecolor=RedViolet,urlcolor=BlueGreen,linkcolor=RedViolet]{hyperref}
\usepackage[normalem]{ulem}
\usepackage{url}
\usepackage{graphicx,wrapfig,float,slashed,cancel}
\usepackage{amsmath,amssymb,epsfig,xcolor,stmaryrd}
\usepackage{bm}
\usepackage{enumitem}
\usepackage{hhline,multirow,tabularx} 
\usepackage{graphics}
\usepackage{latexsym}
\usepackage{rotating}
\usepackage{subfigure}
\usepackage{color}
\usepackage{blindtext}
\usepackage{lipsum}
\usepackage{multirow}
\usepackage{cleveref}
\usepackage{orcidlink}

\DeclareRobustCommand*\cal{\@fontswitch\relax\mathcal}

\allowdisplaybreaks
\begin{document}
	
	
	\title{\textcolor{BlueViolet}{Z boson production in proton-lead collisions: A study utilizing MRW nuclear TMDs}}

	\author{S.~Rezaie$^{a}$\orcidlink{0000-0002-1852-1619}}
	\email{somayeh@ipm.ir }
 
	\author{K.~Azizi$^{b,c,a}$\orcidlink{0000-0003-3741-2167}}
	\email{kazem.azizi@ut.ac.ir}

	\affiliation{
	$^{a}$School of Particles and Accelerators, Institute for Research in Fundamental Sciences (IPM) P. O. Box      19395-5531, Tehran, Iran\\
		$^{b}$Department of Physics, University of Tehran, North Karegar Avenue, Tehran 14395-547, Iran\\
			$^{c}$Department of Physics,  Dogus University, Dudullu-\"{U}mraniye, 34775
		Istanbul, T\"{u}rkiye\\		
	}

	\date{\today}
	
	\preprint{}
	
	\begin{abstract}
	This paper investigates the production of Z bosons in proton-lead collisions at a center of mass energy of $\sqrt{s} = 8.16$ TeV, utilizing various transverse momentum dependent parton distribution functions (TMDs), including the leading order Martin-Ryskin-Watt (LO-MRW), next-to-leading order MRW (NLO-MRW), and parton branching (PB) approaches. By comparing theoretical predictions with experimental data from the CMS collaboration, we assess the performance of these TMD models across different kinematic regions. Our analysis reveals that while both LO-MRW and NLO-MRW models generally align well with experimental data, the LO-MRW model tends to overestimate in certain kinematic regions. The NLO-MRW model, with its strong ordering constraint, provides better agreement in these areas. This study highlights the impact of different impositions of angular and strong ordering constraints in the LO-MRW and NLO-MRW approaches in describing Z boson production.

	\end{abstract}
	
	
	\maketitle
	
	\renewcommand{\thefootnote}{\#\arabic{footnote}}
	\setcounter{footnote}{0}
	
	\section {Introduction}
	The calculation of cross sections in hadronic collisions is inherently challenging due to the nonperturbative nature of the hadrons' internal structure. Factorization theory provides a robust framework to address this complexity by separating processes occurring at different energy scales. Specifically, the proton-proton cross section can be expressed as a convolution of a perturbative partonic cross section and nonperturbative parton distribution functions (PDFs). Two principal approaches in this framework, collinear factorization and $k_t$-factorization, are extensively used for computing cross sections, particularly in lepton-proton and proton-(anti)proton collisions.

In the collinear factorization approach, partons are assumed to move collinearly with their parent hadrons, carrying a specific longitudinal momentum fraction $x$. In this approach, the parton distribution functions used to calculate the cross section depend on the longitudinal momentum fraction $x$ and the factorization scale $ \mu $. These parton distribution functions are referred to as collinear PDFs. Conversely, the $k_t$-factorization approach \cite{KT} incorporates the transverse momentum of partons, accounting for both their intrinsic motion within hadrons and their perturbative evolution from low to high energy scales. This leads to the formulation of transverse momentum dependent distributions (TMDs), also known as unintegrated PDFs (UPDFs).

At high energies, where the longitudinal momentum fraction $x$ becomes small, the role of transverse momentum becomes increasingly significant. Consequently, incorporating transverse momentum into PDFs and partonic cross sections is crucial for realistic modeling of parton dynamics. Unlike the collinear framework, which relies on the Dokshitzer–Gribov–Lipatov–Altarelli–Parisi (DGLAP) evolution equations \cite{DokshitzerDeepInelastic, altarelli, gribovDeepInelastic}, the $k_t$-factorization approach employs evolution equations tailored for TMDs. Historically, TMD evolution has been described using the BFKL (valid at small $x$) \cite{BFKL1, BFKL2, BFKL3, BFKL4} and CCFM (applicable across small and large $x$) \cite{CCFM1, CCFM2, CCFM3, CCFM4} equations. However, these methods have predominantly addressed gluon distributions, creating a need for approaches that encompass both quarks and gluons. To bridge this gap, DGLAP evolution equations have been adapted to derive TMDs for both quarks and gluons, as implemented in methods such as Kimber-Martin-Ryskin (KMR) \cite{KMR}, Martin-Ryskin-Watt (MRW) \cite{MRW}, and Parton Branching (PB) \cite{PB1, PB2}. These methods have been successfully used to calculate various proton-(anti)proton cross sections and exhibit strong agreement with experimental data \cite{WattWZ, BermudezMartinez:2019anj, SomZ, D, kord_validity}.

The production of electroweak bosons, particularly Z boson, play a pivotal role in exploring parton dynamics and testing the validity of theoretical models in particle physics. Z boson, due to their weak interactions with the strong QCD medium, serves as a clean probe of the initial state, providing critical insights into PDFs, TMDs, and the partonic content of hadrons \cite{SomZ, D, kord_validity}. At small transverse momentum, the Z boson's cross section is particularly sensitive to TMDs, offering a unique opportunity to validate various factorization approaches and evolution equations.

Recent advances in hadronic structure have highlighted the importance of transverse momentum factorization in hadronic-nuclear collisions \cite{lipatov, hanes, Bury}. Proton-proton interactions serve as a baseline for understanding hadronic processes, given their well characterized PDFs. However, extending this framework to proton-lead collisions introduces additional complexities due to nuclear effects. These effects necessitate the use of nuclear PDFs (nPDFs), which account for the influence of additional nucleons in the scattering process. The transition from proton-proton to proton-lead interactions provides a valuable perspective on how nuclear environments impact hadronic structure and particle dynamics. Notably, the study of proton-lead interactions is crucial for developing theoretical models that describe nuclear-nuclear collisions, which are fundamental to modern and future collider experiments.

An intriguing example is the investigation of Z boson production cross sections in proton-lead collisions using nuclear TMDs. Recent studies \cite{hanes} utilized the PB approach, implemented in TMDLib \cite{TMDLIB}, alongside the KaTie parton level event generator \cite{KATIE}, to compute the Z boson production cross section at $\sqrt{s} = 5.02$ TeV \cite{CMS5.02}. In this work, we extend these efforts to explore Z boson production at $\sqrt{s} = 8.16$ TeV \cite{CMS8.16} in proton-lead collisions. Using the KaTie event generator, we analyze both experimental data and theoretical frameworks. Unlike prior investigations, we incorporate a detailed comparison of the LO-MRW and NLO-MRW TMD models in addition to the PB approach. Although the LO-MRW and NLO-MRW formalisms have been applied to Z boson production in proton-proton collisions \cite{SomZ, kord_validity, D}, their use in proton-lead collisions has been limited. Notably, while some studies \cite{lipatov, Bury} have explored the LO-MRW approach in this context, they are either lack of experimental comparisons or focus on heavy-flavor production using alternative modeling techniques. Our work thus represents the first comprehensive analysis of both LO-MRW and NLO-MRW TMDs for Z boson production in proton-lead collisions, offering insights into the MRW approach's applicability to nuclear TMDs (nTMDs).

The structure of this paper is as follows: Section II outlines the $k_t$-factorization framework and the KaTie parton level event generator. Section III introduces the nTMDs used in this study. Section IV discusses the numerical methods employed.  Section V presents our results and discussion. Finally, section VI summarizes our findings and conclusions.

\section {$k_t$-factorization framework}
\label{sec:kt-factorization}

The $k_t$-factorization framework provides a systematic approach for calculating cross sections in high energy hadronic collisions, particularly for processes like Drell-Yan lepton pair production. In this study, we employ the KaTie parton level event generator to compute cross sections within this framework. The cross section can be expressed as:

\begin{equation}
    \label{eq:kt-cs}
    \sigma = \sum_{a,b=q,g}\int \frac{dx_{1}}{x_{1}} \frac{dx_{2}}{x_{2}} \frac{dk^2_{1t}}{k^2_{1t}} \frac{dk^2_{2t}}{k^2_{2t}} f_{a}(x_{1}, k^2_{1t}, \mu^2) f_{b}(x_{2}, k^2_{2t}, \mu^2) \hat{\sigma}^*_{ab}.
\end{equation}

In this equation, $\hat{\sigma}^*_{ab}$ represents the off-shell partonic cross section for partons $a$ and $b$, which accounts for their transverse momentum. The functions $f_{a}(x_{1}, k^2_{1t}, \mu^2)$ and $f_{b}(x_{2}, k^2_{2t}, \mu^2)$ are the TMDs for partons $a$ and $b$, respectively.

The TMDs, $f_{a(b)}(x, k_t^2, \mu^2)$, depend on three key parameters: the fractional longitudinal momentum $x$ of the parton relative to its parent hadron, the transverse momentum $k_t$ of the parton, and the factorization scale $\mu$ at which the TMDs are evaluated.

The KaTie parton level event generator, implemented primarily in Fortran with Python components, facilitates the calculation of hadronic cross sections as defined in \cref{eq:kt-cs}. The generator conveniently accepts input files that specify the relevant subprocesses and experimental kinematic constraints. For incorporating TMDs, users can either utilize the TMDLib package, which provides access to a predefined set of TMDs, or supply custom TMD grid files containing columns for $x$, $k_t^2$, $\mu^2$, and $f(x, k_t^2, \mu^2)$ for each parton flavor.

KaTie performs the necessary interpolations to obtain the required TMDs for calculating the hadronic cross section. In this work, we utilize TMDLib for the PB TMDs, while for LO-MRW and NLO-MRW TMDs, we provide custom grid files to the KaTie event generator. The specific TMDs employed in this study are detailed in the following section.
	
\section {The nuclear TMDs }	
\subsection{PB TMDs}
\label{sec:2b}

The PB approach provides a systematic mechanism for evolving parton densities from a low initial scale, where the distributions are parameterized, to the relevant hard process scale. The evolution is governed by the DGLAP evolution equations, solved iteratively to account for each individual parton splitting process, while respecting kinematic constraints. By associating a physical interpretation with the evolution scale, the PB approach allows for the calculation of the transverse momentum of partons during evolution, unlike the MRW method, which only accounts the transverse momentum at the final step of evolution( for a detailed discussion, see \cite{NuclB}). This leads to the construction of TMDs \cite{PB2}.

In this method, the evolution is carried out using NLO DGLAP splitting functions, with angular ordering applied. The evolution scale $\mu_i$ is connected to the transverse momentum of the emitted parton through the relation:
\begin{equation}
	{\bf q}_{t,i}^2 = (1-z_i)^2\mu_i^2,
\end{equation}
where $z_i$ is the momentum fraction of the splitting. The nuclear TMD, ${ F}^{\mathrm{Pb}}_a(x, k_t^2, \mu^2)$, is expressed as a convolution of an initial TMD distribution and an evolution kernel:
\begin{equation}
	x{F}^{Pb}_a(x, k_t^2, \mu^2) = \int dx' \, {F}^{\mathrm{Pb}}_{0,b} (x',  k_{t,0}^2, \mu_0^2) 
	\frac{x}{x'} \, {K}_{ba}\left(\frac{x}{x'}, k_{t,0}^2, k_t^2, \mu_0^2, \mu^2\right),
\end{equation}
where $F^{\mathrm{Pb}}_{0,b}(x', k_{t,0}^2, \mu_0^2)$ represents the initial TMD distribution, and ${K}_{ba}$ is the evolution kernel. For simplicity, the intrinsic transverse momentum distribution of the initial TMD is modeled using a Gaussian function:
\begin{equation}
	{F}^{\mathrm{Pb}}_{0,b}(x, k_{t,0}^2, \mu_0^2) = f^{Pb}_{0,b}(x, \mu_0^2) \cdot \exp\left(-{| k_{t,0}^2|}/{\sigma^2}\right),
\end{equation}
with $\sigma^2 = q_0^2 / 2$ and $q_0 = 0.5 \, \text{GeV}$, and $f^{Pb}_{0,b}(x, \mu_0^2)$ are the PDFs at the initial scale.

In this work, we utilize the predictions of the reference \cite{hanes}, in which they use the PB-nCTEQ15FullNuc\_208\_82-set2 TMD set, where is based on the nCTEQ15 nuclear PDF, and is also prefered choice as it is shown in the same reference. This set uses an initial scale of $\mu_0^2 = 1.4 \, \text{GeV}^2$ and defines the strong coupling $\alpha_s$ using the transverse momentum of the splitting process. This choice effectively incorporates coherence effects and higher order corrections.

\subsection{LO-MRW and NLO-MRW TMDs}
\label{LO-NLO-MRWSec}
LO-MRW and NLO-MRW TMDs are additional DGLAP-based TMDs, alongside PB, that enable the determination of nuclear TMDs for both quarks and gluons. The general form of TMDs based on the MRW method for the (anti) quarks and gluon is given by \cite{MRW}:
\begin{equation}
	\label{eq:MRW}
	f_a(x, k_t^2, \mu^2) = T_a(k_t^2, \mu^2) \dfrac{\alpha_s(k_t^2)}{2 \pi} \sum_{b=q,g} \int_x^1 dz P_{ab}(z) f_b(\dfrac{x}{z}, k_t^2 )\Theta^{\delta^{ab}}(z_{max} - z),
\end{equation}
where $T_a(k_t^2, \mu^2)$ represents the Sudakov form factor (FF), $f_b(\frac{x}{z}, k_t^2)$ indicates the momentum weighted parton densities at the leading order (LO) level, and $\Theta^{\delta^{ab}}(z_{max} - z)$ enforces a limitation on gluon emission when $a=b$.

The Sudakov FF, can be expressed in the following manner: 
$$
T_a(k_t^2\leq\mu^2, \mu^2) = exp\left(-\int_{k_t^2}^{\mu^2}  \dfrac{d\kappa_t^2}{\kappa_t^2} \dfrac{\alpha_s(\kappa_t^2)}{2 \pi} \sum_{b=q,g} \int_0^1 d\xi \xi P_{ba}(\xi)\Theta^{\delta^{ab}}(\xi_{max} - \xi) \right) ,
$$
\begin{equation}
	\label{eq:Sud}
	T_a(k_t^2 > \mu^2, \mu^2) = 1.
\end{equation}
The MRW TMDs are derived from the DGLAP evolution equations, enabling the calculation of TMDs for both quarks and gluons. In such method, it is considered  that partons evolve collinearly along the parent proton until the final evolution step, following the DGLAP  equation. Just  at the final  step, through a real emission, the partons acquire $k_t$ dependence. Subsequently, the partons evolve to the factorization scale $\mu$ without any real emission, governed by the Sudakov FF.
	
It is important to note that \cref{eq:MRW} is applicable only at $k_t \geq \mu_0$, where $\mu_0 \sim 1$ GeV, since input PDFs, $f_a(x, k_t^2)$, are not expressed  at scales below $\mu_0$. To extend the MRW TMDs to $k_t < \mu_0$, one can use the normalization limitation \cite{MRW, SomZ, kord_inclusive_jet}:

\begin{equation}
\label{eq:Norm}
f_a(x, \mu^2) = \int_{0}^{\mu^2} \dfrac{dk_t^2}{k_t^2} f_{a}(x, k_t^2, \mu^2),
\end{equation}
which ensures the TMDs satisfy the normalization condition. For $k_t < \mu_0$, the following distribution fulfills this condition:
\begin{equation}
\label{eq:MRWLessMu0}
\dfrac{1}{k_t^2}f_a(x, k_t^2<\mu_0^2, \mu^2) = \dfrac{1}{\mu_0^2} f_a(x,\mu_0^2 )T_a(\mu_0^2,\mu^2).
\end{equation}
This leads to a constant distribution at $k_t < \mu_0$.

Finally, the LO-MRW TMDs for (anti) quarks and gluons, along with their respective Sudakov form factors, are expressed as follows \cite{kord_inclusive_jet, Kord3photon, Olanj}:

\begin{equation}
	\label{eq:LOMRWQ}
	\begin{split}
			f_q(x,k_t^2, \mu^2) = T_q(k_t^2, \mu^2) \frac{\alpha_{s}^{LO}(k_t^2)}{2\pi}\int_x^1\Big[ P_{qq}^{LO}(z)f_q^{LO}(\frac{x}{z}, k_t^2)\Theta(z_{max}-z)  \\
			+ P_{qg}^{LO}(z) f_g^{LO}(\frac{x}{z}, k_t^{2})\Big]\;\mathrm{d}z,
		\end{split}
\end{equation}
\begin{equation}
	\label{eq:LOMRWG}
	\begin{split}
			f_g(x,k_t^2, \mu^2) = T_g(k_t^2, \mu^2) \frac{\alpha_{s}^{LO}(k_t^2)}{2\pi}\int_x^1\Big[  P_{gg}^{LO}(z)f_g^{LO}(\frac{x}{z}, k_t^2)\Theta(z_{max}-z)\\
			+ \sum_q P_{gq}^{LO}(z)f_q^{LO}(\frac{x}{z}, k_t^2)\Big]\;\mathrm{d}z,
	\end{split}	
\end{equation}
with the Sudakov FFs given by:
\begin{equation}
	\label{eq:8}
	T_q(k_t^2,\mu^2) = exp\left(-\int_{k_t^2}^{\mu^2} \frac{\mathrm{d}\kappa_t^2}{\kappa_t^2} \frac{\alpha_{s}^{LO}(\kappa_t^2)}{2 \pi}\int_{0}^1 P_{qq}^{LO}(\xi)\Theta(\xi_{max} - \xi)\;\mathrm{d}\xi \right),	
\end{equation}
\begin{equation}
	\label{eq:9}
	T_g(k_t^2,\mu^2) = exp\left(-\int_{k_t^{2}}^{\mu^2}\frac{\mathrm{d}\kappa_t^2}{\kappa_t^2} \frac{\alpha_{s}^{LO}(\kappa_t^2)}{2 \pi}\int_{0}^1 \left[ \xi P_{gg}^{LO}(\xi)\Theta(\xi_{max}-\xi) \Theta(\xi - \xi_{min})
	+ n_F  P_{qg}^{LO}(\xi)\right] \;\mathrm{d}\xi \right).
\end{equation} 
In the equation above, $q$ represents quarks ${u,\overline{u}, d, \overline{d},\ldots}$, and $n_F$ represents the number of active quark flavors.
To set the cutoff for $z$ and $\xi$ in order to constrain gluon emissions in  \cref{eq:MRW,eq:Sud}) one can adopt either the strong ordering or angular ordering approach in the final step of evolution. The strong ordering cutoff (SOC) serves as an approximation of the angular ordering cutoff (AOC) when $z$ approaches the large limit, i.e., $z\to 1$ \cite{KIM1}. Consequently, if the primary interest lies in effects at small  $x$, the AOC is the more suitable choice. The AOC cutoff is given by:

\begin{equation}
	q_{i+1} > z_i q_{i}, 
\end{equation}
where $q=k_t/(1-z)$, and in the final stage  of the evolution, this cutoff can be expressed  as:
\begin{equation}
	\mu > z_{max} k_t / (1-z)\to z_{max} = \mu / (\mu + k_t).
\end{equation}
In the case of  SOC, i.e., the $z\to1$ limit, this  cutoff can be represented as:
\begin{equation}
	\mu > k_t / (1-z_{max})\to z_{max} = 1 - k_t/\mu.
\end{equation}	
It is clear that the SOC cutoff is more restrictive, confining the parton's transverse momentum to the region where $k_t < \mu$, while the AOC is less stringent, permitting the parton transverse momentum to exceed the factorization scale. As noted in reference \cite{KIM2}, the KMR approach was initially formulated using the SOC.  But, as indicated in references \cite{KMR, KIM3}, the SOC cutoff has  been supplanted by the more precise AOC. Consequently, the MRW method exclusively employed the AOC \cite{MRW}. Therefore, based on the preceding discussion, we also endorse the AOC in the present study for the MRW UPDFs, which is given by: $z_{max} = \frac{\mu}{(\mu + k_t)}$ and $\xi_{max} = \frac{\mu}{(\mu + \kappa_t)}$.

The LO-MRW formalism is extended  to NLO by modifying the DGLAP scale and employing NLO splitting functions, as shown in \cref{eq:NLOMRWQ,eq:NLOMRWG}. This formalism is developed to the NLO-level by the choice of the DGLAP scale as $k^2 = k_t^2/(1 - z)$ instead of $k_t^2$. However, Martin et al.~\cite{MRW} demonstrated that using LO splitting functions produces similar TMDs, thereby simplifying the computation. Furthermore, an additional SOC, $\Theta(\mu^2 - k^2)$, is introduced to constrain the parton distribution to ${k_t}^2 < \mu^2$. This SOC significantly impacts the behavior of TMDs at large $k_t$ ($k_t \simeq \mu$) and large $z$ ($z \simeq 1$), leading to their suppression in these regions.

In contrast, the corresponding LO-MRW formalism lacks such constraints, resulting in larger TMDs compared to NLO-MRW. Notably, it is not restricted to $k_t \leq \mu$. Martin et al.~\cite{MRW} also showed that the use of NLO splitting functions in this formalism has minimal effect on the resulting TMDs, and similar outcomes can be achieved by using LO splitting functions.

In this work, we adopt this simplified approach; henceforth, the NLO-MRW formalism is expressed as follows~\cite{MRW, Kord3photon, kord_inclusive_jet}:
\begin{equation}
\label{eq:NLOMRWQ}
\begin{split}
f_q(x,k_t^2, \mu^2) =\int_x^1 T_q(k^2, \mu^2) \frac{\alpha_{s}^{NLO}(k^2)}{2\pi} \Big[ P_{qq}^{LO}(z)f_q^{NLO}(\frac{x}{z}, k^2)\Theta(z_{max}-z)  \\
+ P_{qg}^{LO}(z) f_g^{NLO}(\frac{x}{z}, k^{2})\Big]\Theta(\mu^2-k^2) \;\mathrm{d}z,
\end{split}
\end{equation}
	
\begin{equation}
\label{eq:NLOMRWG}
\begin{split}
f_g(x,k_t^2, \mu^2) = \int_x^1 T_g(k^2, \mu^2) \frac{\alpha_{s}^{NLO}(k^2)}{2\pi} \Big[  P_{gg}^{LO}(z)f_g^{NLO}(\frac{x}{z}, k^2)\Theta(z_{max}-z)\\
+ \sum_q P_{gq}^{LO}(z)f_q^{NLO}(\frac{x}{z}, k^2)\Big] \Theta(\mu^2-k^2)\;\mathrm{d}z,
\end{split}	
\end{equation}
	
with sudakov FFs as:
\begin{equation}
\label{eq:12}
T_q(k^2,\mu^2) = exp\left(-\int_{k^2}^{\mu^2}  \frac{\mathrm{d}p^2}{p^2} \frac{\alpha_{s}^{NLO}(p^2)}{2 \pi}\; \int_{0}^{1} \;\mathrm{d}\xi \; P_{qq}^{LO}(\xi) \Theta(\xi_{max}-\xi) \right),
\end{equation}
	
\begin{equation}
\label{eq:13}
T_g(k^2,\mu^2) = exp\left(-\int_{k^2}^{\mu^2}  \frac{\mathrm{d}p^2}{p^2} \frac{\alpha_{s}^{NLO}(p^2)}{2 \pi}\; \int_{0}^{1} \;\mathrm{d}\xi \; \left[P_{gg}^{LO}(\xi) \Theta(\xi_{max}-\xi) \Theta(\xi-\xi_{min}) + n_F P_{qg}^{LO}(\xi)\right] \right).
\end{equation}
	
Due to the dependence of $k^2$ on $z$, the strong coupling constant and Sudakov FFs in \cref{eq:NLOMRWQ,eq:NLOMRWG} must be included within the $z$ integral. This dependency makes the integration process in the NLO-MRW formalism significantly more complex and computationally intensive. We also consider AOC in the present study for the NLO-MRW UPDFs.

The MRW formalism  has also an alternative derivative based representation \cite{MRW}, in addition to the integral version discussed here. However, as highlighted in \cite{ValeshabadiOnTheAmbiguity, Golec_on_KMR, Guiot_pathology}, the derivative form can produce negative and discontinuous TMDs, making it unsuitable for accurate calculations. To ensure correct TMDs, the integral version is necessary.

Before concluding this section, it is important to address a significant point regarding the dimensionality of the LO-MRW method. In its original formulation, the LO-MRW has no inherent dimension, whereas some other sets of UPDFs found in the literature possess a dimension of $ \text{1/(GeV}^2) $. This discrepancy leads to variations in the expression for hadronic cross sections. Specifically, the term $ {1}/{k_t^2} $ in the denominator of \cref{eq:kt-cs} can be effectively incorporated into the UPDFs. As a result, we can express the cross section formula, i.e. \cref{eq:kt-cs}, using $F(x, k_t^2, \mu^2) = \frac{f(x, k_t^2, \mu^2)}{k_t^2}$, as:
\begin{equation}
\sigma = \sum_{a,b =q, g} \int  \frac{dx_1}{x_1} \frac{dx_2}{x_2} dk_{1t}^2 \, dk_{2t}^2 \, F_a(x_1, k_{1t}^2, \mu^2) F_b(x_2, k_{2t}^2, \mu^2) \hat{\sigma}^*_{ij}.	
\end{equation}

\section {numerical methods}
To calculate proton-lead collisions for the CMS experimental data \cite{CMS8.16}, we use different TMD models, specifically the MRW approach at both the LO and NLO levels. For cross section calculations, we generally use the KaTie parton level event generator. However, because the NLO-MRW is not available in TMDLib, and the LO-MRW implementation in the TMDLib package follows the problematic derivative form, as discussed in \cref{LO-NLO-MRWSec}, we generate custom grid files using the MRW formalism for cross section calculations with KaTie. In generating these grid files, we use CT18NLO PDFs \cite{T} as inputs for the LO-MRW ($f_{q(g)}^{LO}$ in \cref{eq:LOMRWQ,eq:LOMRWG}) and NLO-MRW formalisms ($f_{q(g)}^{NLO}$ in \cref{eq:NLOMRWQ,eq:NLOMRWG}) for protons, and nCTEQ15FullNuc\_208\_82 PDF sets for lead, utilizing the LHAPDF C++ library package \cite{B}. Due to the unavailability of LO PDF sets for the nuclear state, we are compelled to utilize NLO PDF sets for both the proton and nuclear components. It is important to note that the use of NLO PDFs in the MRW method has been explored in other studies, such as reference \cite{Marcin}. In this work, our focus is to investigate the effects of the AOC and SOC constraints imposed on the LO-MRW and NLO-MRW approaches within the MRW method.

To evaluate  the $Z$ boson production for LO and NLO-MRW TMDs in $p\text{Pb}$ collisions, we consider the subprocesses $q +\overline{q} \rightarrow Z \rightarrow \ell^+ + \ell^-$ and $q+g \rightarrow Z + q \rightarrow \ell^+ + \ell^- + q$ with five partonic flavors at a center of mass energy of $\sqrt{s} = 8.16 \, \text{TeV}$. While for the PB TMDs, we do not perform the calculations ourselves; instead, we use the predictions provided in reference \cite{hanes}.

We carry out these parton level cross section computations for the first five quark flavors, i.e., up, down, strange, charm, bottom, including their respective antiquarks. For the factorization and renormalization scales, we adopt $\mu_f = \mu_r = \sqrt{p_t^{ll\;2} + m^{ll\;2}}$, where $p_t^{ll}$ and $m^{ll}$ are the transverse momentum and the invariant mass of the output dilepton, respectively. It should be noted that the KaTie parton level event generator calculates the results in the center of mass frame; hence, one needs to boost the results to the lab frame to compare them with the experimental data. We perform this boost for the LO and NLO-MRW approaches on the raw files of generated events from the KaTie parton level generator using our custom C++ code, rather than the histogram tools available in the KaTie parton level event generator.
\begin{figure}
 	\centering
 	\includegraphics[width=8.1cm, height=7.1cm]{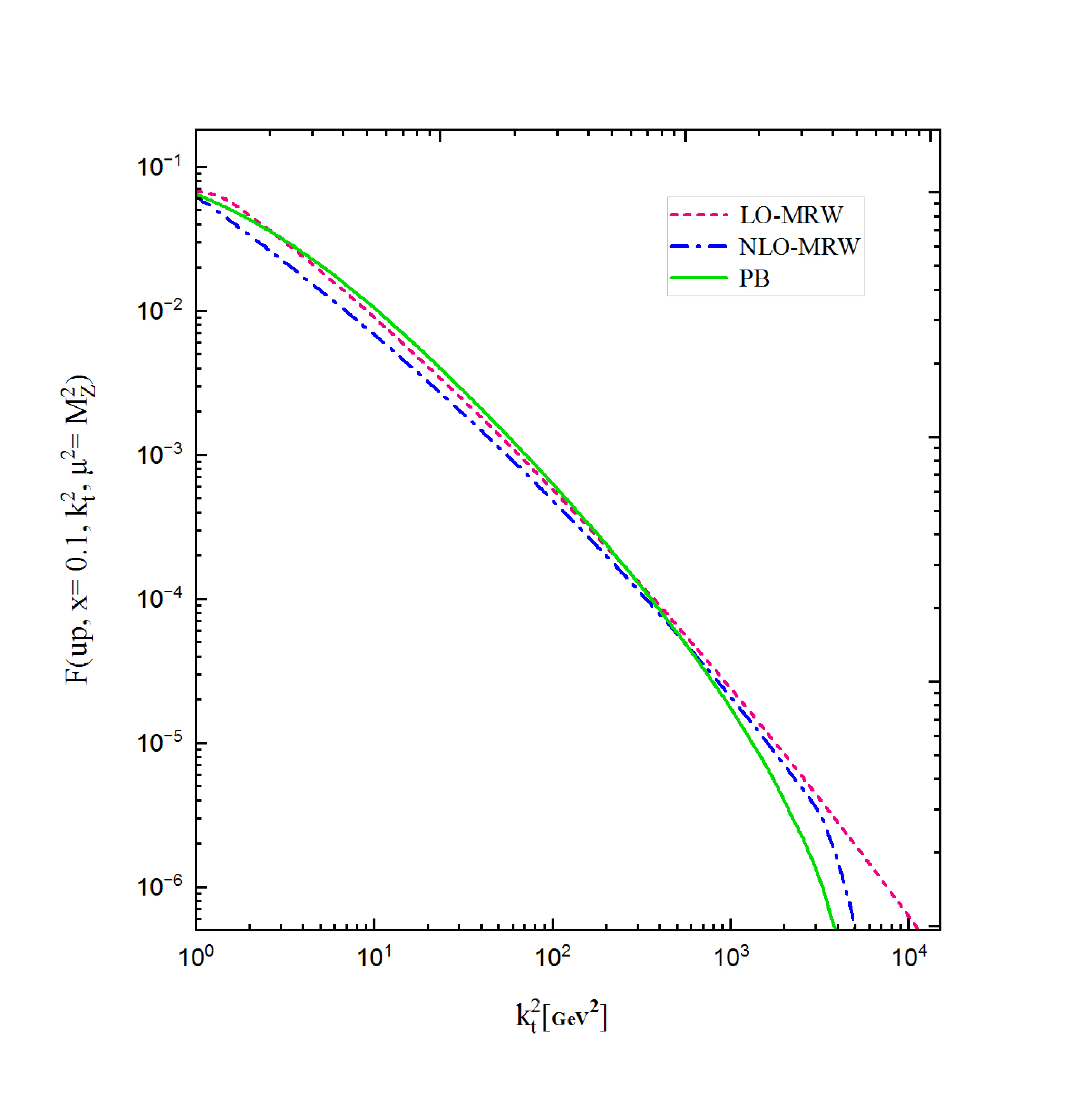}
 	\includegraphics[width=8.1cm, height=7.1cm]{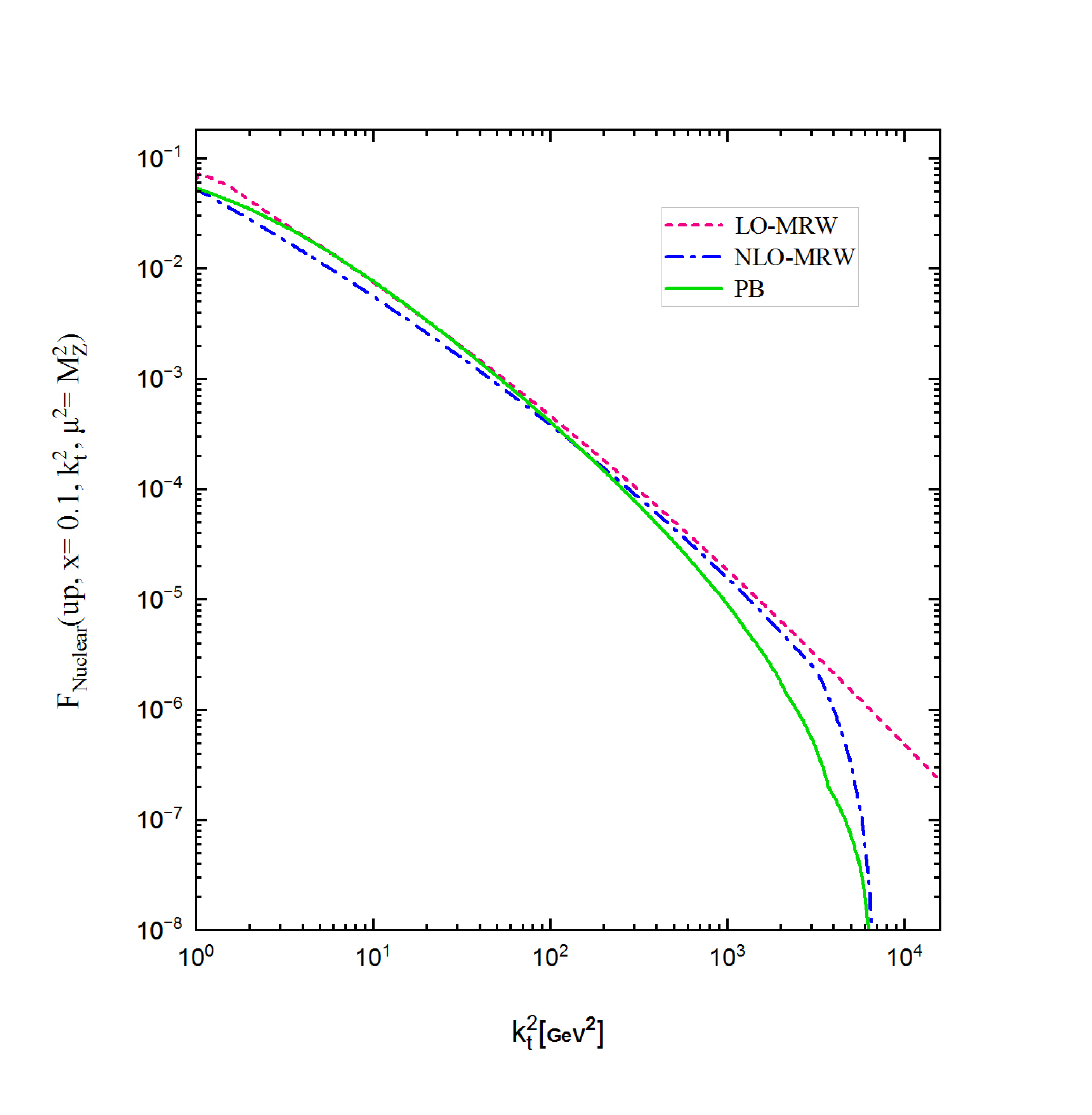}
 	\includegraphics[width=8.1cm, height=7.1cm]{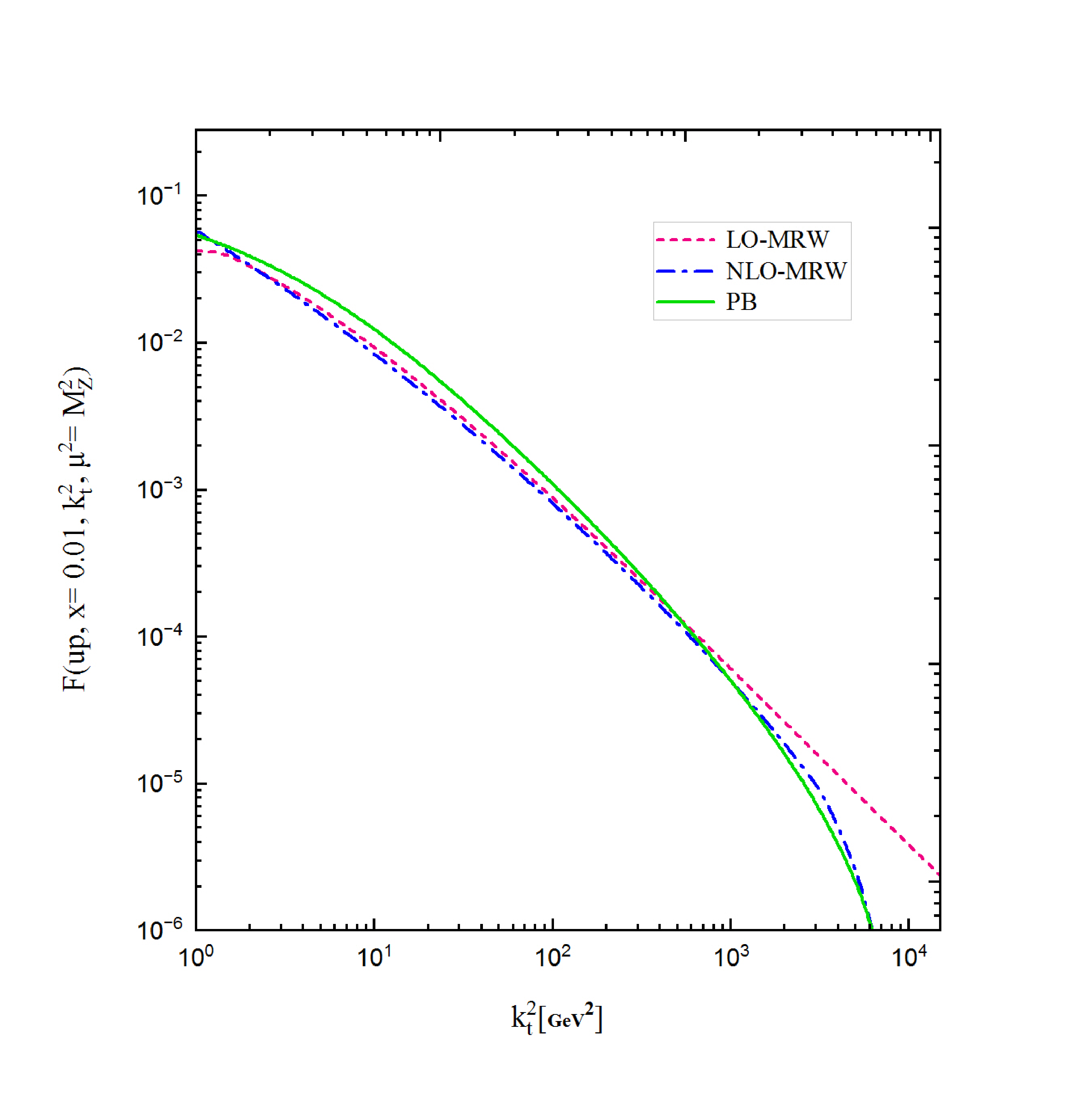}
 	\includegraphics[width=8.1cm, height=7.1cm]{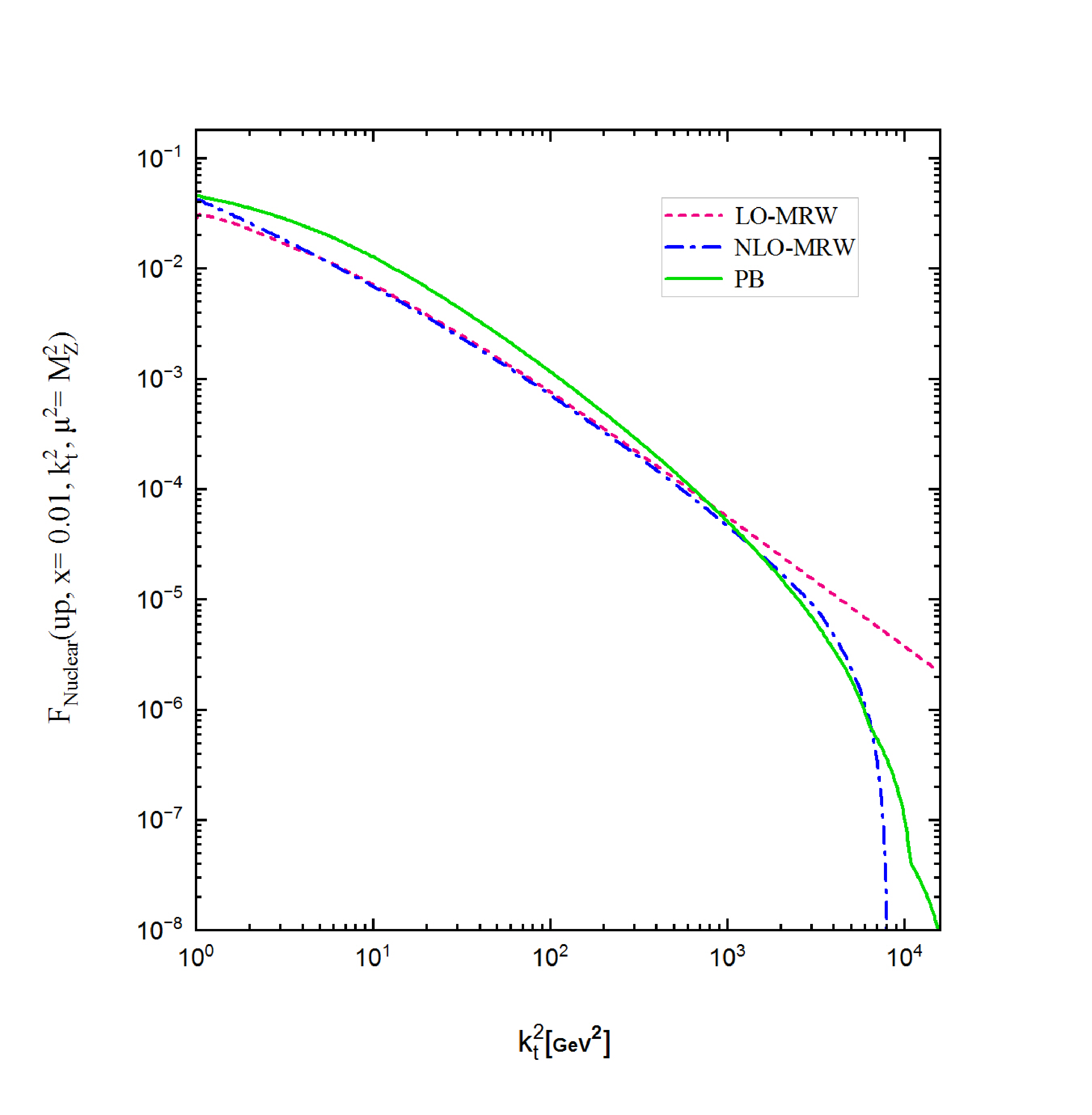}
 	\includegraphics[width=8.1cm, height=7.1cm]{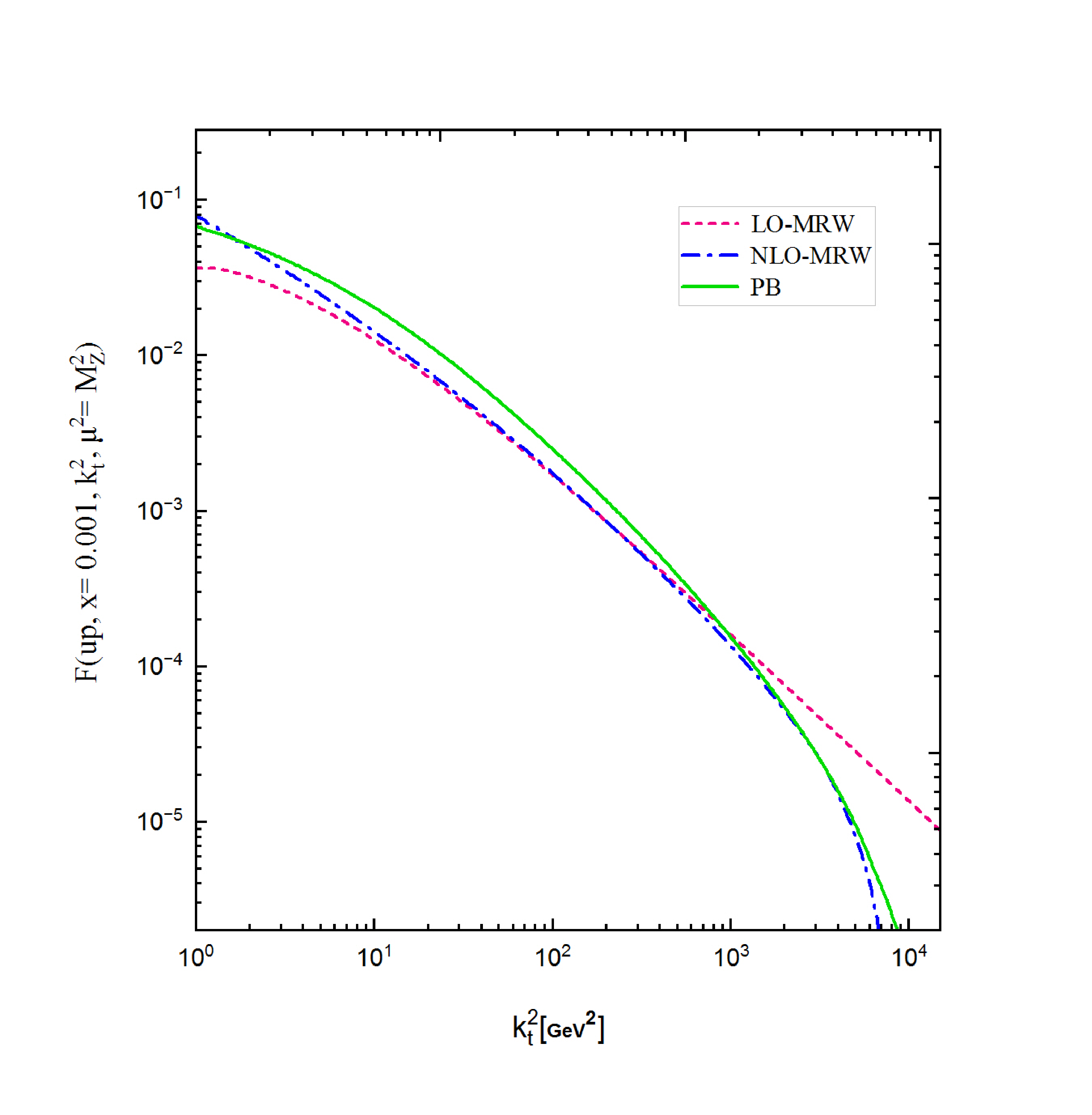}	
 	\includegraphics[width=8.1cm, height=7.1cm]{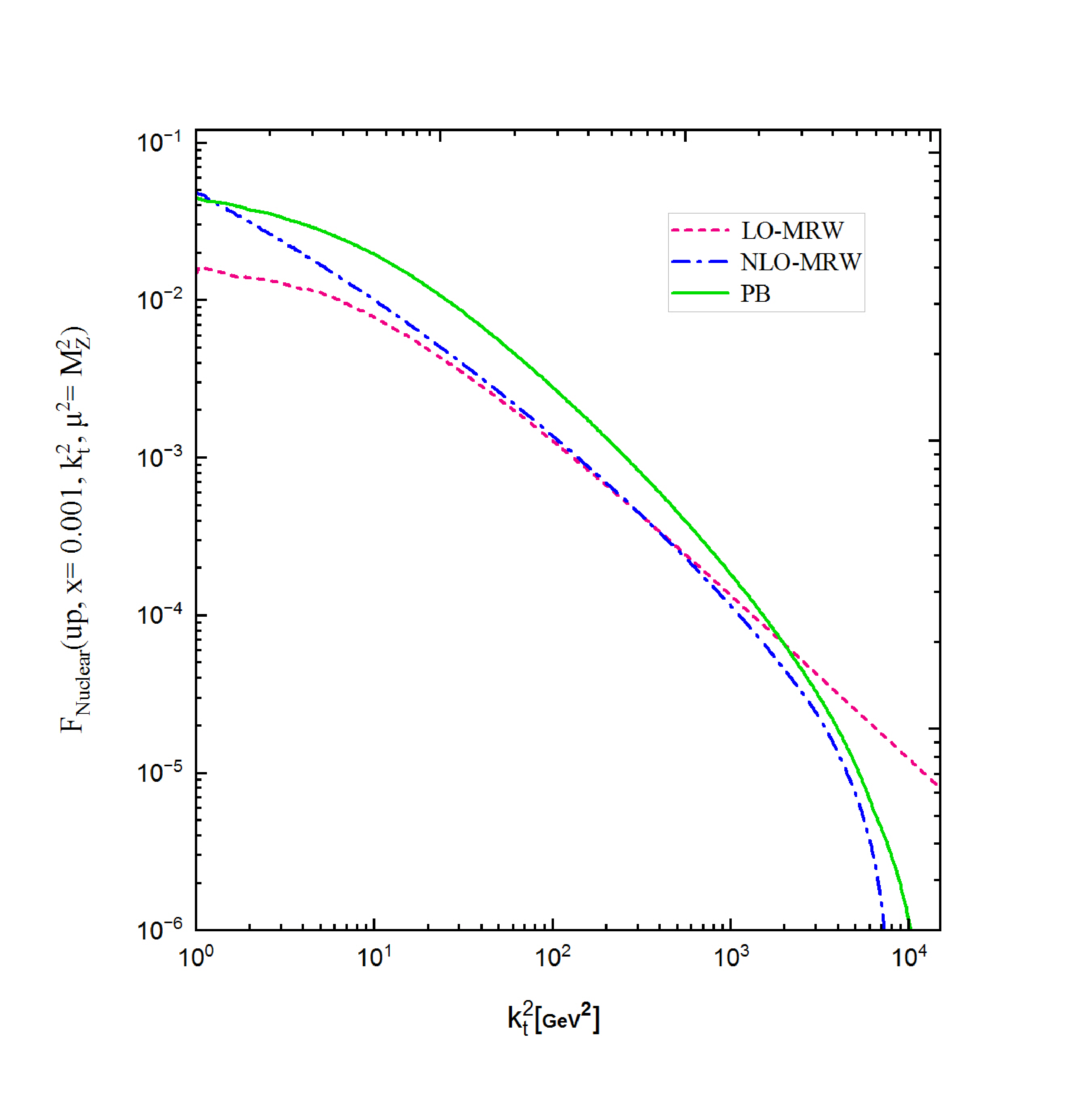}
 	\caption
 	{In the left plots, LO-MRW, NLO-MRW and PB up quark TMDs are shown with respect to $ k_t^2 $ for different $ x $  values, the right  plots are corresponding  nuclear TMDs. }
 	\label{fig:upTMDComparison}
 \end{figure}
 \begin{figure}
 	\centering
 	\includegraphics[width=8.1cm, height=7.1cm]{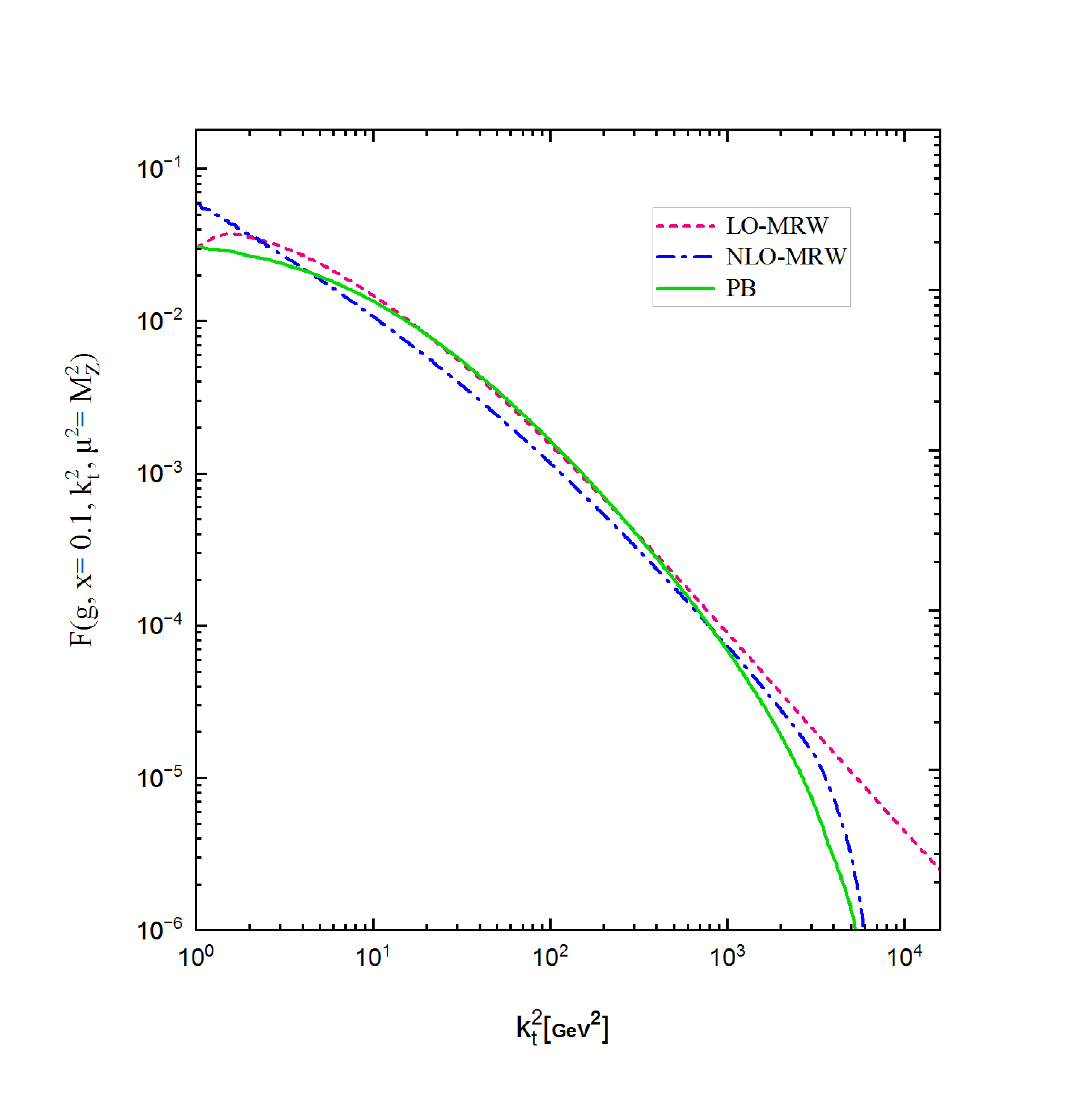}
 	\includegraphics[width=8.1cm, height=7.1cm]{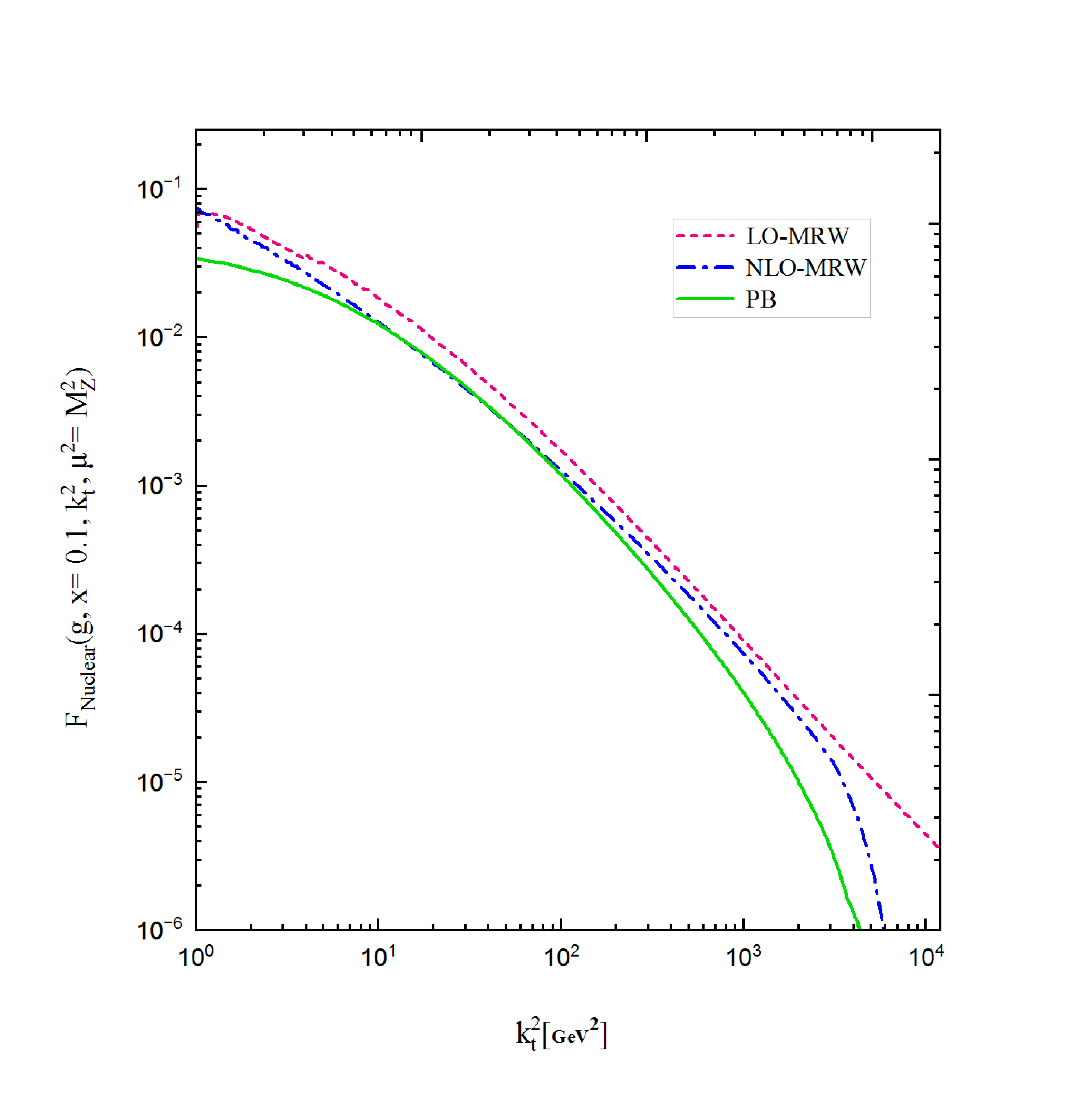}
 	\includegraphics[width=8.1cm, height=7.1cm]{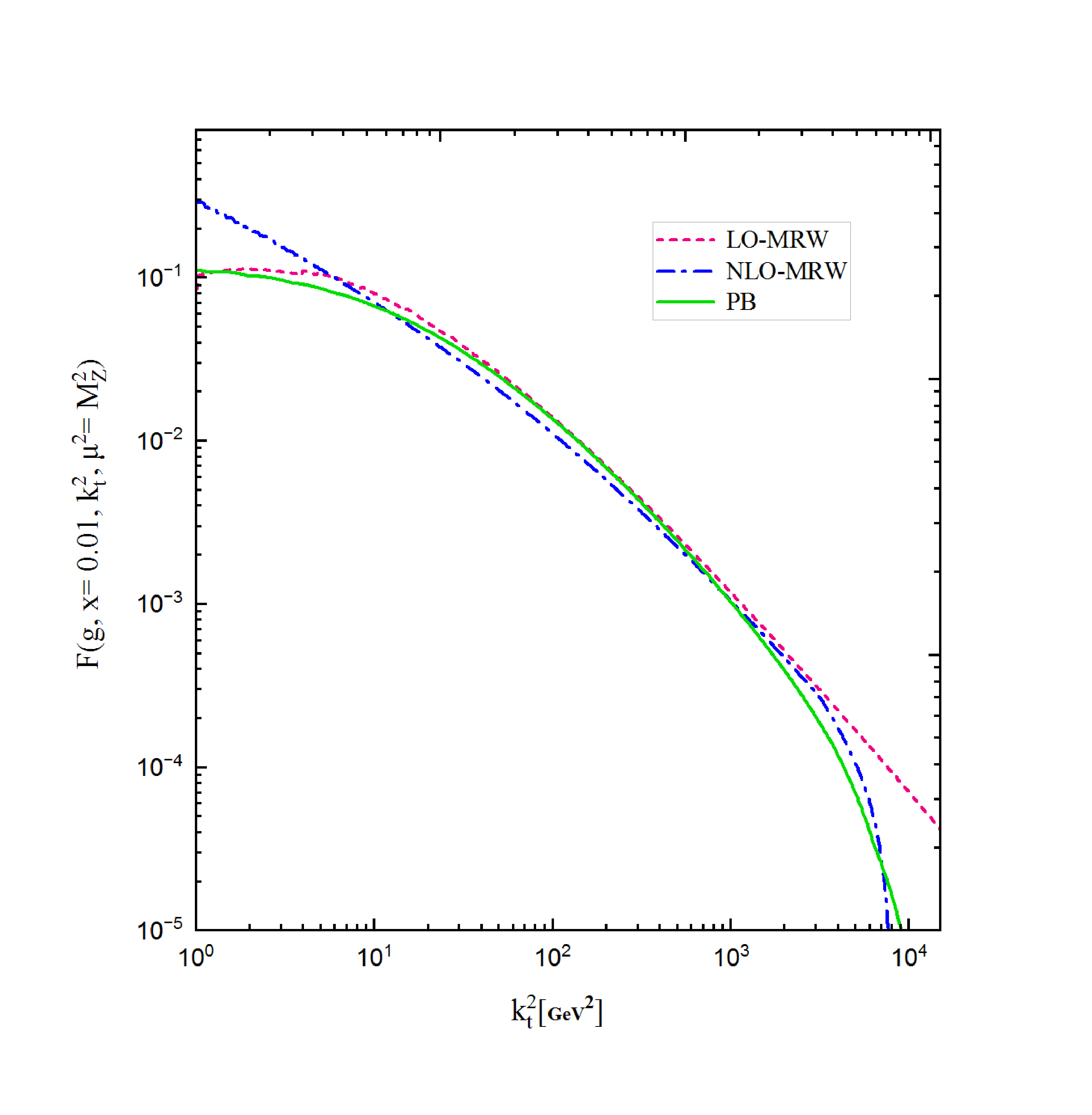}
 	\includegraphics[width=8.1cm, height=7.1cm]{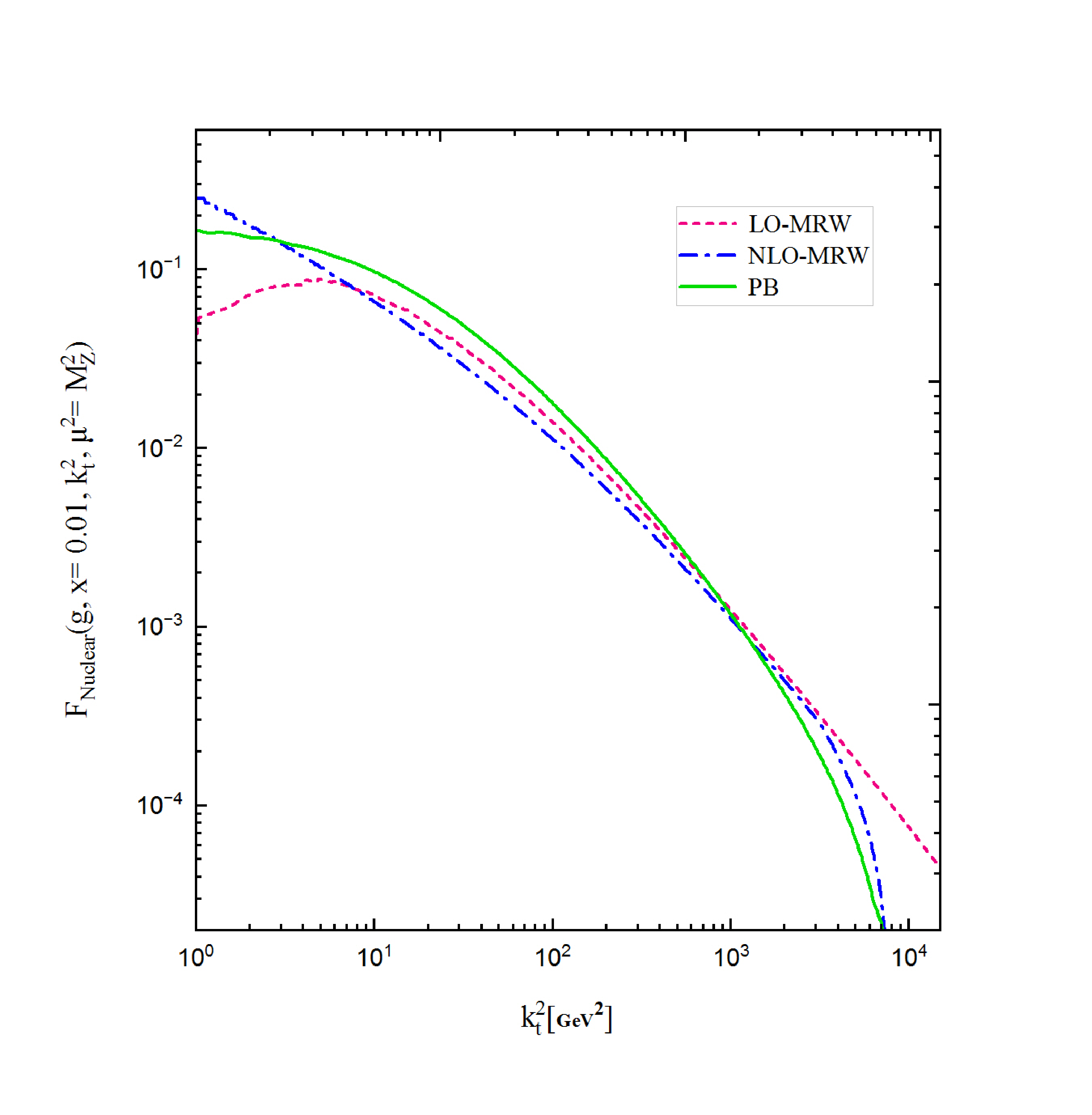}
 	\includegraphics[width=8.1cm, height=7.1cm]{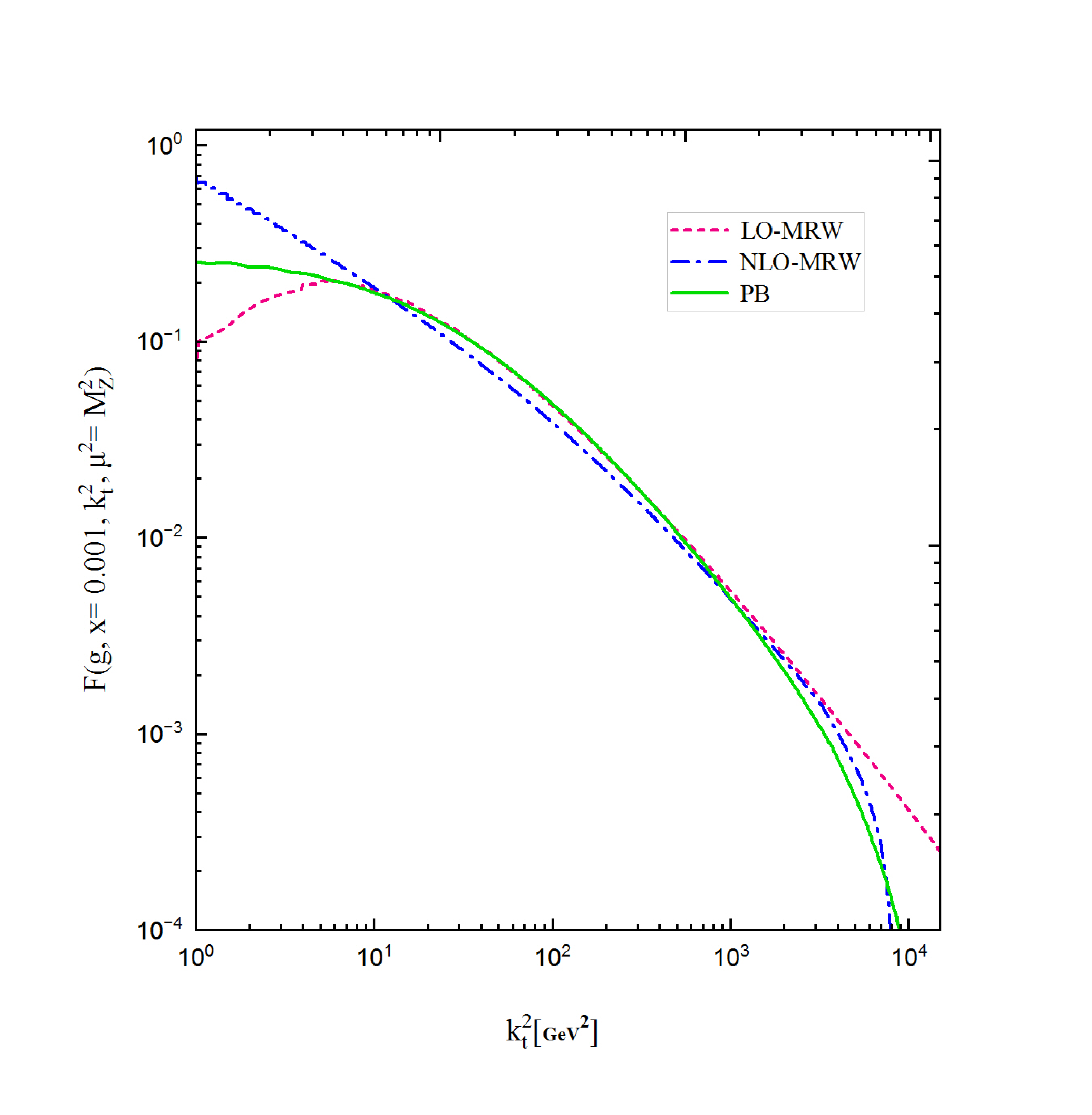}	
 	\includegraphics[width=8.1cm, height=7.1cm]{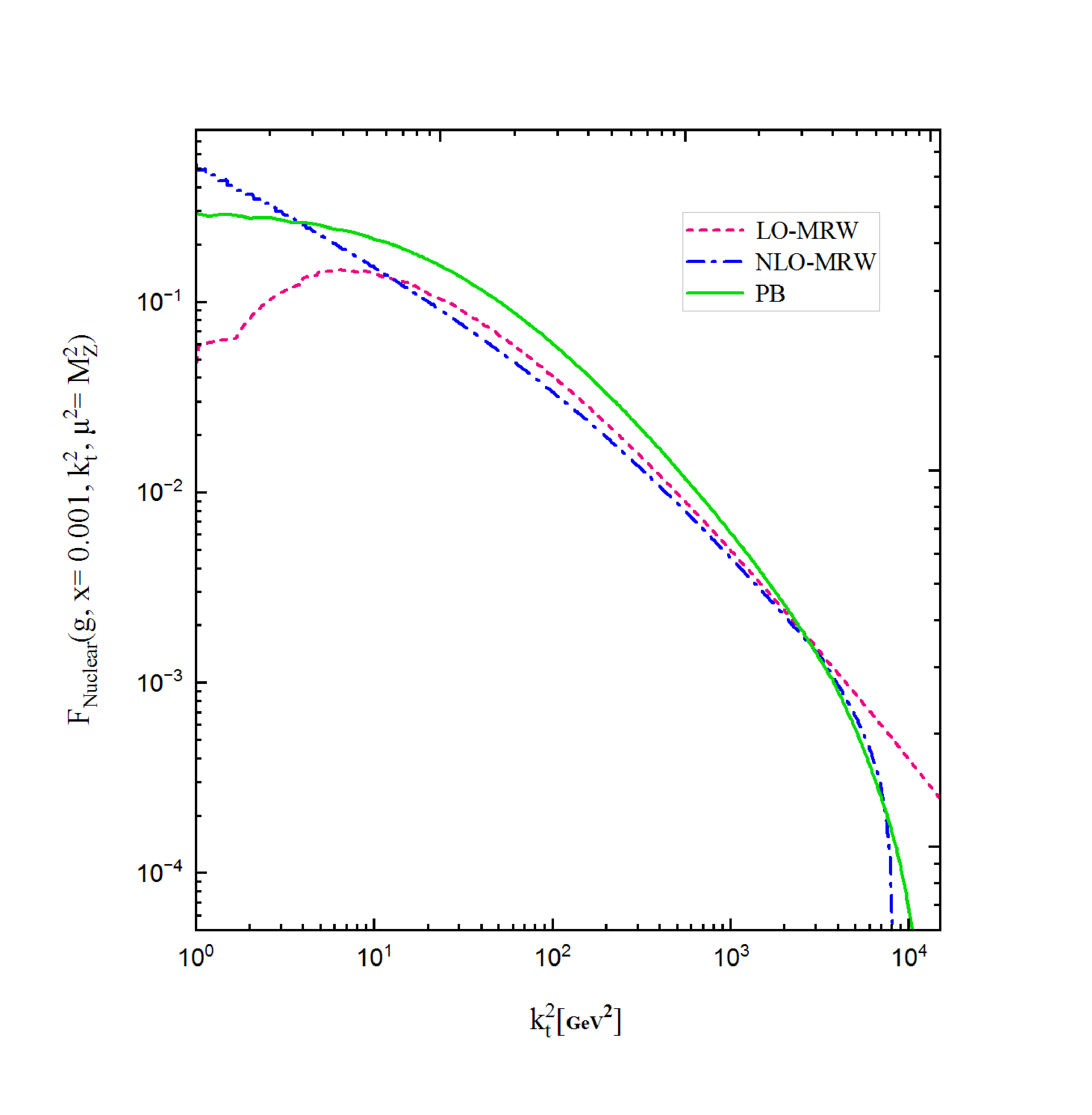}
 	\caption
 	{The left  plots  show LO-MRW, NLO-MRW and PB gluon TMDs with respect to $ k_t^2 $ for different  $ x $ values, the  right  plots depict the corresponding nuclear TMDs.}
 	\label{fig:gTMDComparison}
 \end{figure}
\section {results and discussion}
In this section, we present the results of our calculations and compare them with experimental data from the CMS collaboration at $8.16 \, \text{TeV}$. We calculate differential cross sections for LO-MRW, NLO-MRW TMDs, and PB predictions from the reference \cite{hanes} (by extracting predictions from the figures of this reference). To calculate scale uncertainty, we repeat the event generation with $\mu_{F,R}^{upper} = 2\mu_{F,R}^{central}$ and $\mu_{F,R}^{lower} = 0.5\mu_{F,R}^{central}$. We compare our predictions to the CMS data in the $15 < m_{\mu\mu} < 60 \, \text{GeV}$ and $60  < m_{\mu\mu} < 120 \, \text{GeV}$ regions with respect to the kinematic variables $d\sigma/dP_{t}^{ll}$, $d\sigma/dy_{CM}$, and $d\sigma/d\phi_{\eta}^{*}$. The kinematic variable $\phi_{\eta}^{*}$ is expressed  as:
\begin{equation}
	\label{eq:phistar}
	{\phi^\ast_{\eta}} = \tan(\dfrac{\phi_{acop}}{2})\times \sin(\theta^\ast_\eta).
\end{equation}
In the equation above, $\phi_{acop} = \pi - \vert \Delta \phi \vert$, and $\cos(\theta^\ast_\eta) = \tanh\left[(\eta^- - \eta^+)/2\right]$. Here, $\Delta \phi$ represents the azimuthal angle in the units of  radians between the two leptons, and $\eta^{-(+)}$ denote the pseudorapidity of the negatively (positively) charged leptons. This parameter holds greater  interest to experimentalists than the dilepton transverse momentum because it depends only on the measurement of the final state leptons' directions, alternative to  their precise momenta \cite{SomZ}.

Before presenting the results, we compare different TMD models to gain a deeper understanding of their impact. By examining these models, we can better understand how the choice of input TMD models affects the outcomes of our analysis on the cross section of Drell-Yan lepton pair production.

In \cref{fig:upTMDComparison,fig:gTMDComparison}, we compare various TMD models for the up quark and gluon at $\mu^2 = 10000 \, \text{GeV}^2$ for values of $x = 0.1$, $x = 0.01$, and $x = 0.001$. Generally, the NLO-MRW TMD model's parton distribution at large transverse momentum is more suppressed and is limited to $k_t < \mu$ compared to PB and LO-MRW TMD models. This suppression is due to the $\Theta(\mu^2-k^2)$, strong ordering cutoff, which becomes more pronounced as $x$ increases. Despite this constraint, the LO-MRW model has greater flexibility, with a soft angular ordering limit on the gluon emission term, while there is no limit on quark emission terms. Hence, the LO-MRW TMD model has a larger contribution as the transverse momentum of partons increases, even at $k_t \geq \mu$. However, all up and gluon TMD distributions for protons at medium transverse momentum regions behave similarly. For lead nuclear TMDs, as $x$ decreases, the PB approach shows a larger distribution compared to other TMD models for both gluon and up quark distributions. 
 \begin{figure}
 	\centering
 	\includegraphics[width=8.1cm, height=7.1cm]{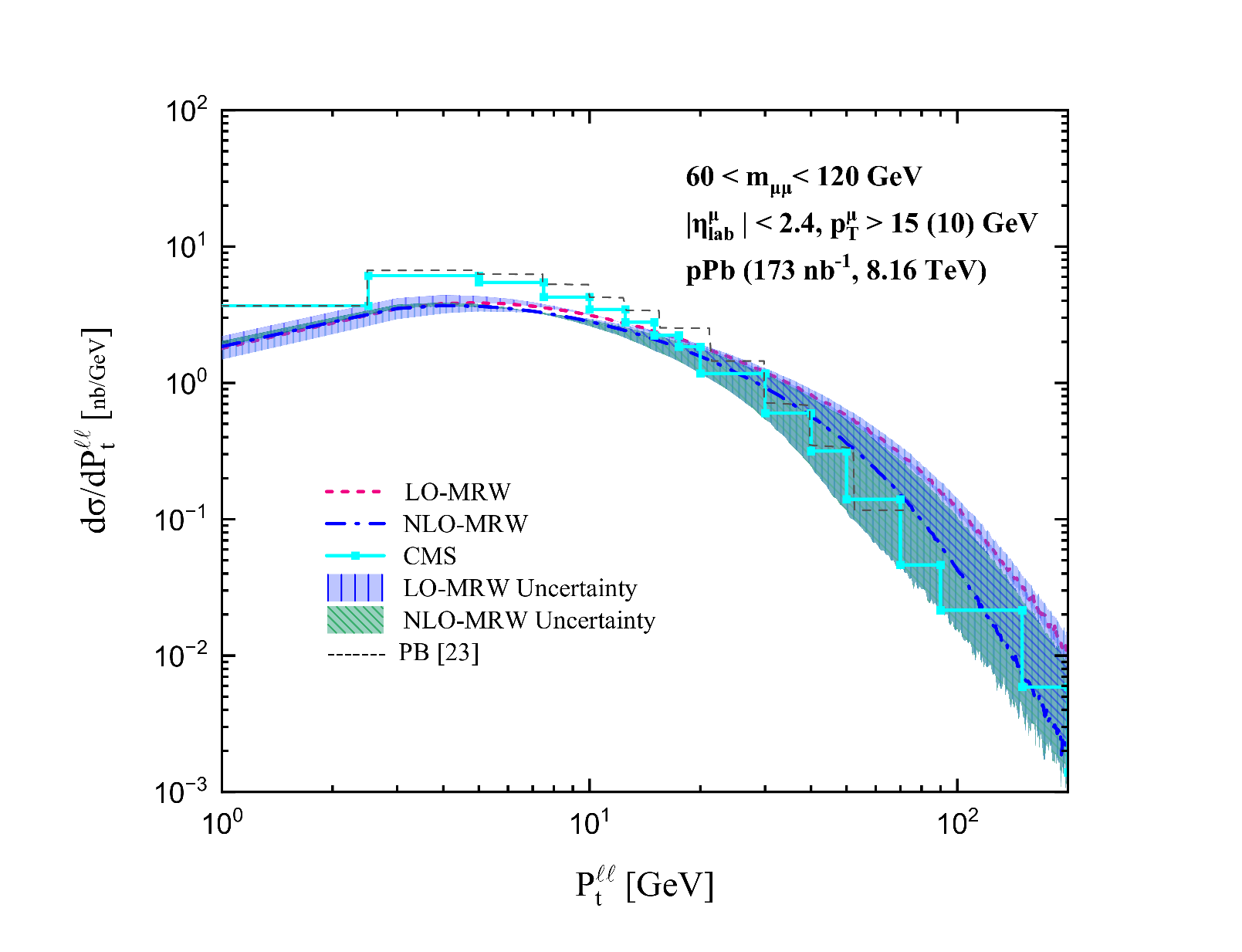}
 	\includegraphics[width=8.1cm, height=7.1cm]{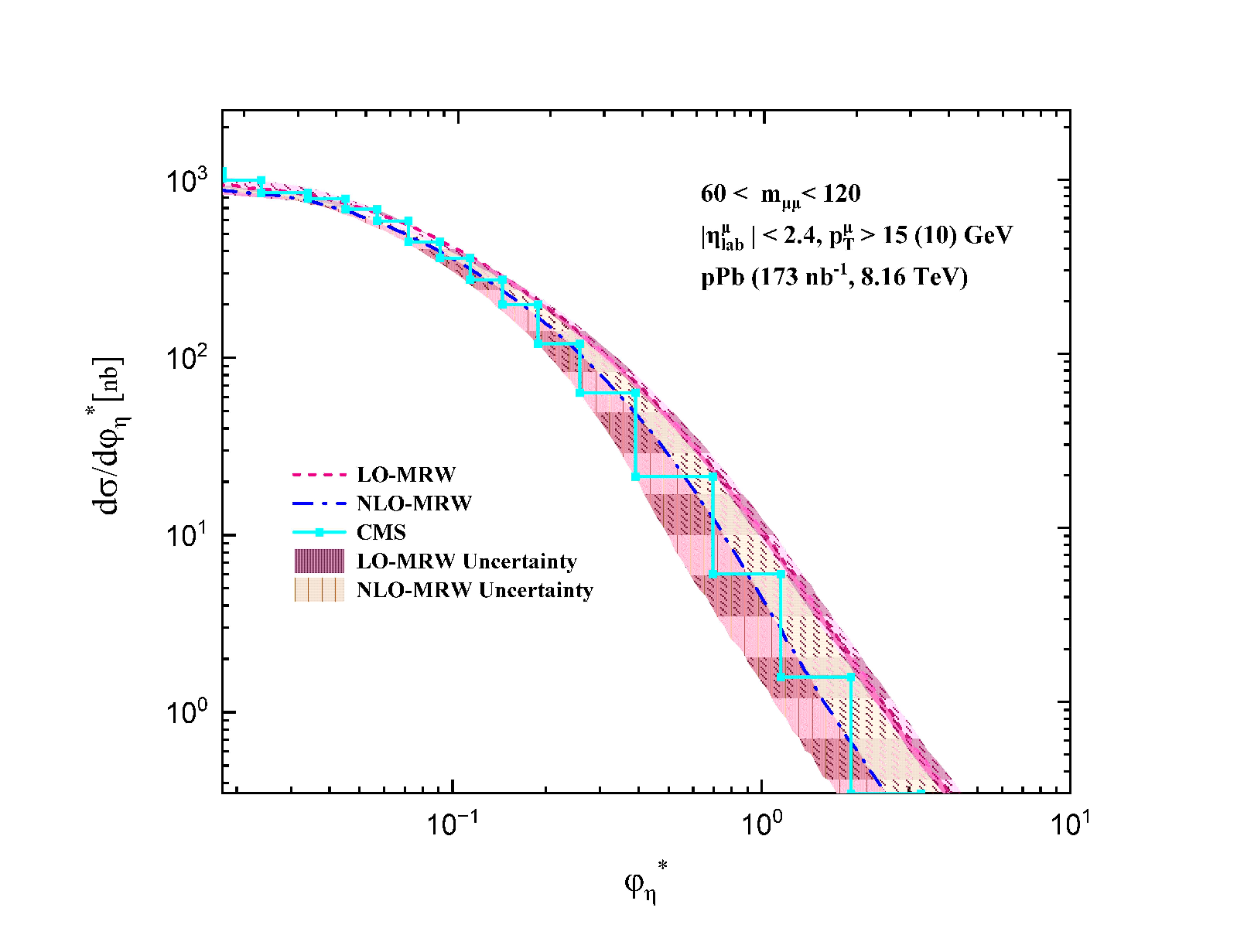}
 	\includegraphics[width=8.4cm, height=7.1cm]{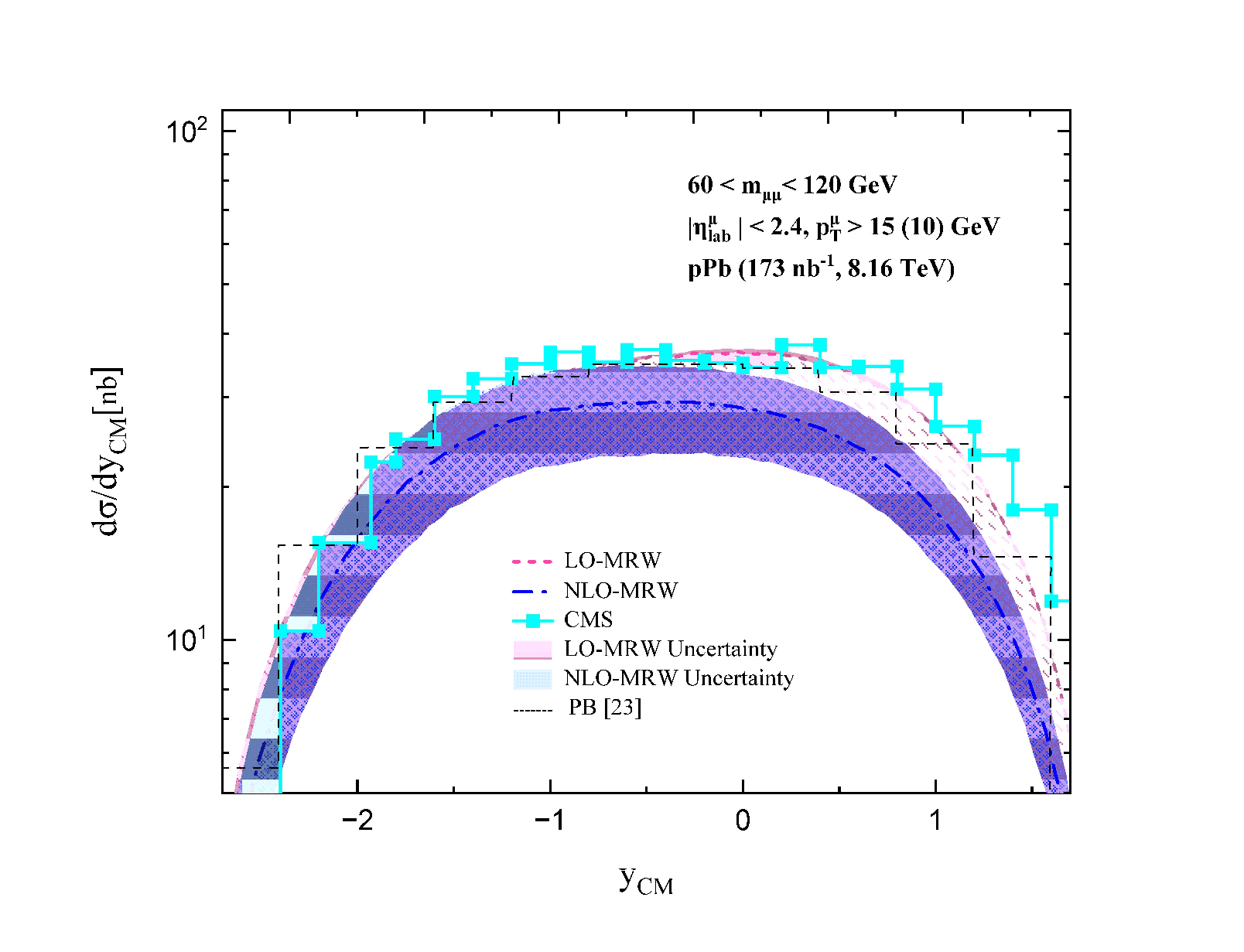}
 	\caption
 	{ Comparison of the differential cross sections with respect to  $P_t^{ll}$, (${\phi}^*_{\eta}$), and  $y_{CM}$ in the  $60 < m_{\mu\mu} < 120\;GeV$ region  using LO-MRW, NLO-MRW and PB TMDs of the reference \cite{hanes}.}
 	\label{fig:All-60-120}
 \end{figure}

 In \cref{fig:All-60-120}, we present a comparison of the differential cross section predictions using LO-MRW, NLO-MRW, and PB TMD models with respect to $P_{t}^{ll}$, ${\phi}^*_{\eta}$, and $y_{CM}$ kinematic variables in the $60 < m_{\mu\mu} < 120 \, \text{GeV}$ region. Note that predictions with PB TMD models are extracted from the figures of the reference \cite{hanes}, so PB predictions' binning and range are not compatible with experimental data, although the experimental constraints are correctly imposed on their cross section predictions. All predictions of different TMD models generally agree well with CMS experimental data. However, in some regions, such as large ${\phi}^*_{\eta}$ and $P_t^{ll}$, the LO-MRW TMD model overestimates experimental data, while the NLO-MRW model aligns better in those regions with the data. This is consistent with our expectations from the comparison of different TMD models in \cref{fig:upTMDComparison,fig:gTMDComparison}, where it was observed that LO-MRW has larger predictions with respect to NLO-MRW and PB distributions at large partonic transverse momenta. Again, according to our expectations from the comparisons, of TMD models, predictions of LO-MRW and NLO-MRW TMD models are similar at medium dilepton transverse momentum, with differences mostly at large partonic transverse momentum. In contrast to differential cross section predictions with respect to $y_{CM}$ using the NLO-MRW TMD model, LO-MRW TMD, and PB have better agreement with this experimental data. However, the NLO-MRW uncertainty band in negative and small center of mass rapidity is close to the experimental data. As $y_{CM}$ increases, corresponding to large $x_A$ (fractional momentum of parton inside proton) limits, the NLO-MRW TMD model is completely suppressed, and even considering scale uncertainty does not improve predictions. NLO-MRW results generally undershoot LO-MRW and PB cross section predictions, due to the strong ordering cutoff, $\Theta(\mu^2-k^2)$, in this model. It should be noted that the PB results generally show better agreement with experimental data. This advantage can be attributed to how the PB approach treats parton emissions along the evolution ladder. In this formalism, the transverse momentum ($k_T$) dependence of parton distribution functions arises from multiple parton emissions. In contrast, the MRW formalism considers only the final step in the evolution \cite{NuclB}. A key implication of this treatment is that, unlike the MRW approach, the PB method allows for control over the structure of the jet associated with Z-boson production. This capability has been successfully applied in studies of Z+jets production using NLO matching \citep{EPJC} and jet merging techniques \cite{JHEP, PLB}. This control over the associated jet structure is crucial for the description of transverse observables like $p_T$ and ${\phi}^*_{\eta}$, which can be further investigated in future works. It should be noted that the data extracted from reference \cite{hanes} do not cover all the regions of our work and a complete comparison with our work in all regions of the transverse momentum of the dilepton is not possible. Therefore, they are presented only for comparison of methods. 
  \begin{figure}
 	\centering
 	\includegraphics[width=8.1cm, height=7.1cm]{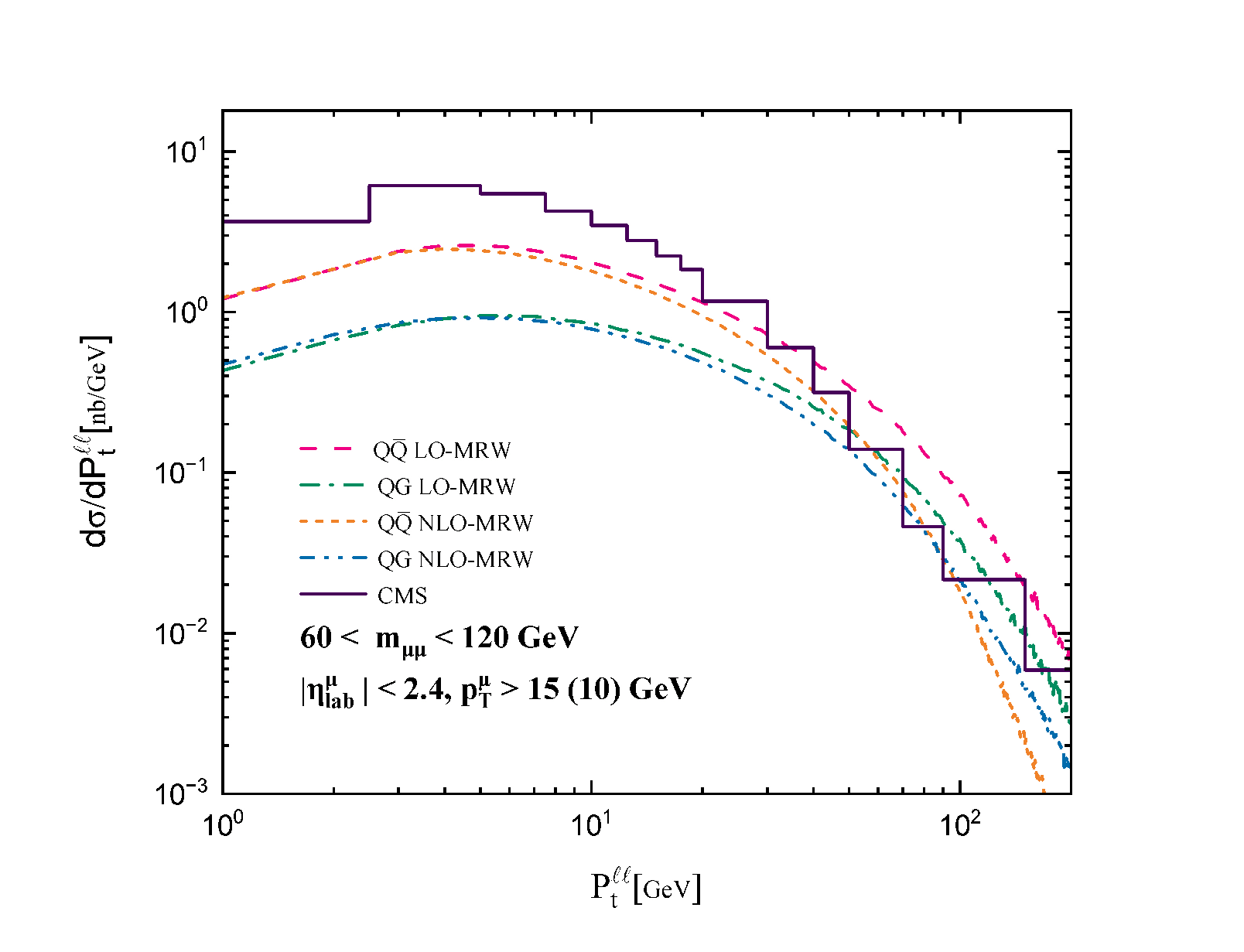}
 	\includegraphics[width=8.1cm, height=7.1cm]{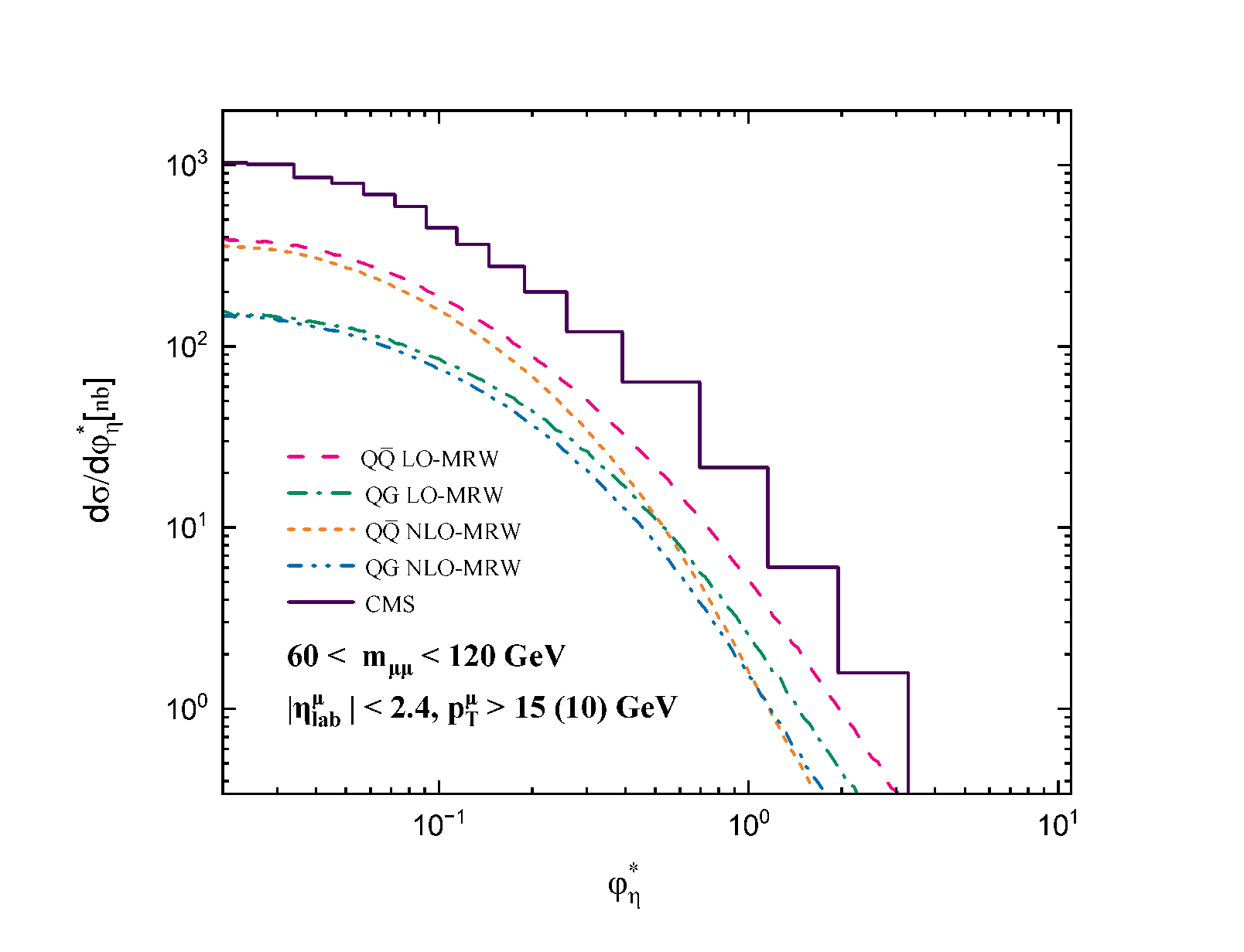}
 	\includegraphics[width=8.4cm, height=7.1cm]{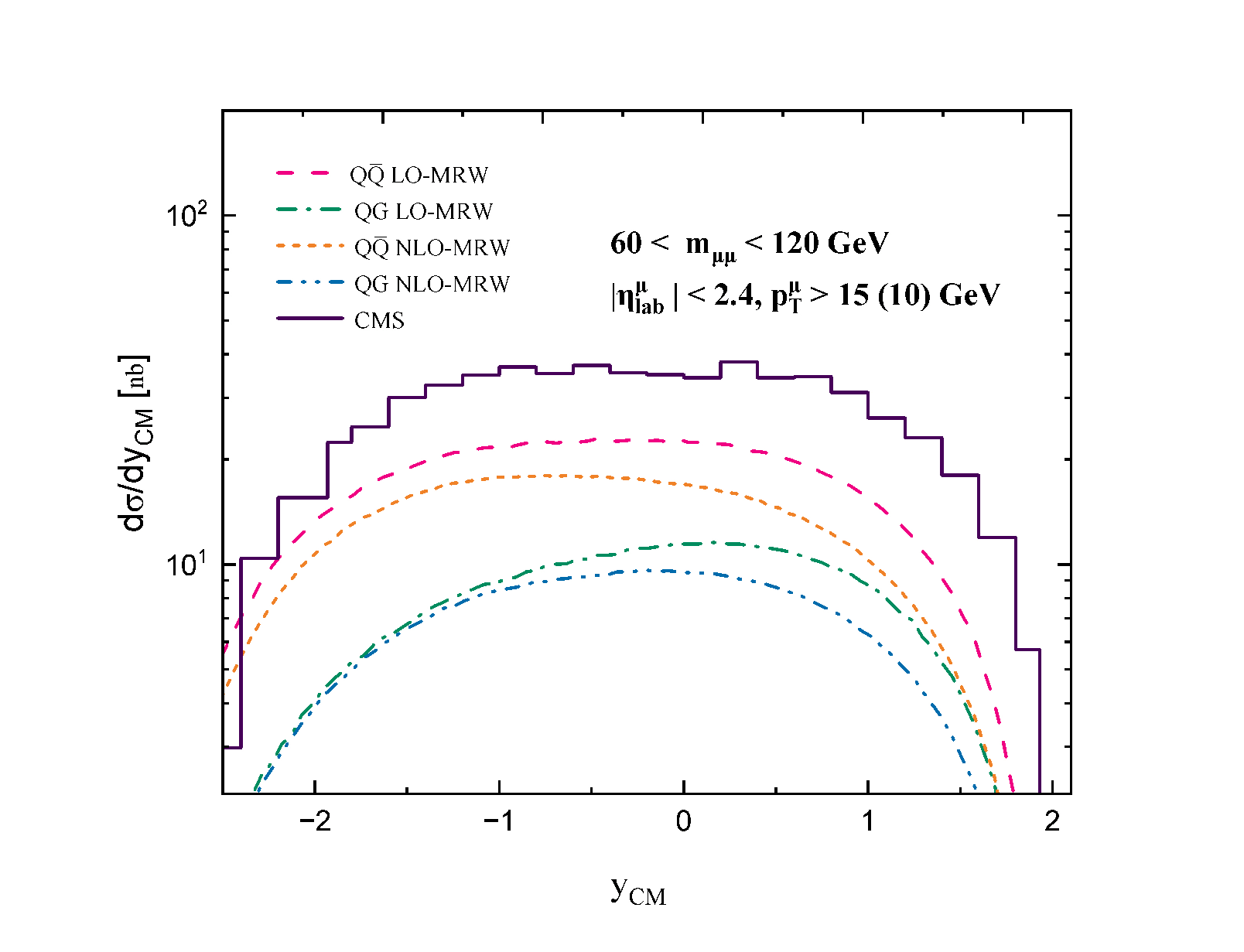}
 	\caption
 	{ Comparison of the differential cross sections of each subprocess with respect to  $P_t^{ll}$, (${\phi}^*_{\eta}$), and  $y_{CM}$ in the  $60 < m_{\mu\mu} < 120\;GeV$ region using LO-MRW, and NLO-MRW TMDs.}
 	\label{fig:Sub-60-120}
 \end{figure}
\begin{figure}
	\centering
	\includegraphics[width=8.1cm, height=7.1cm]{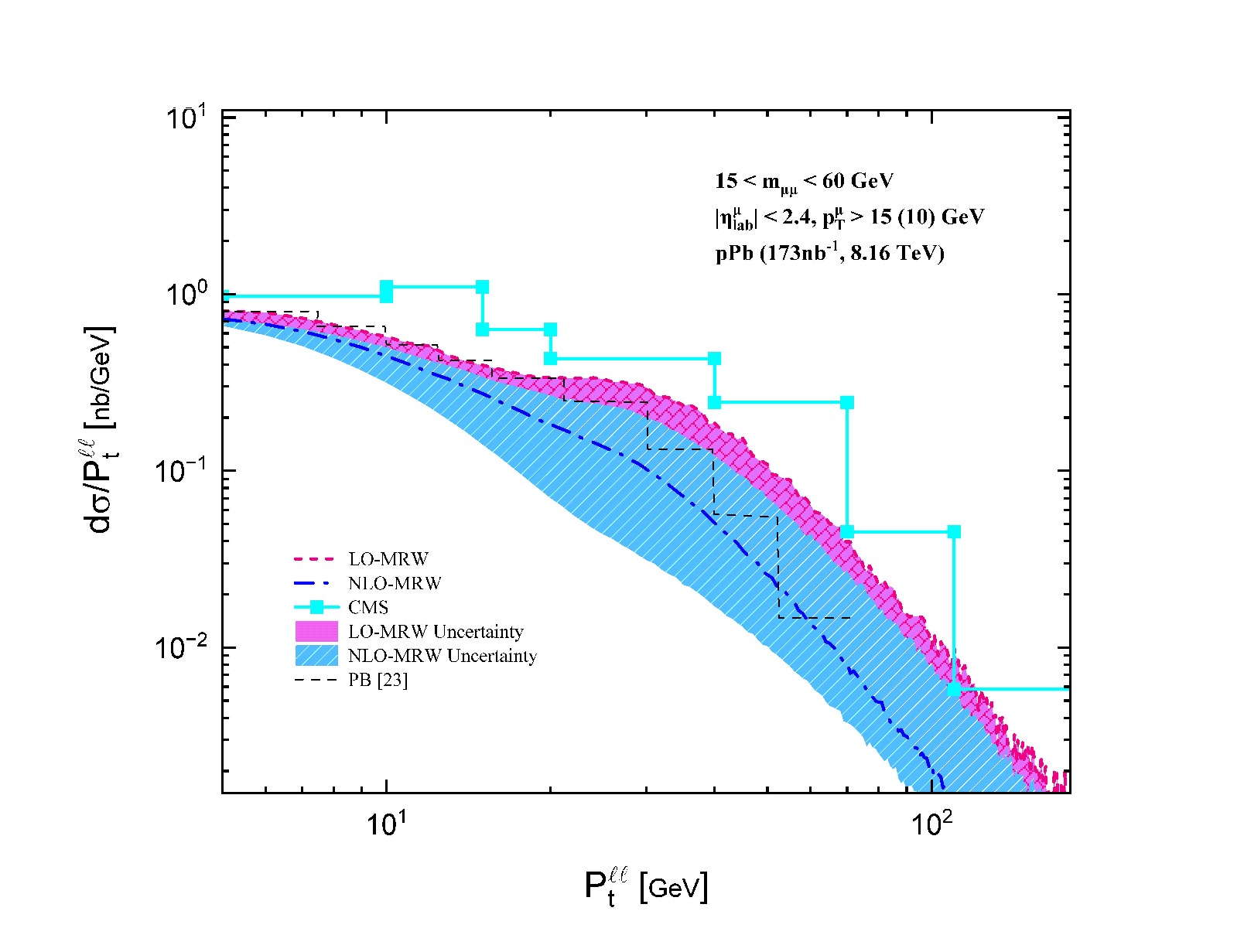}
	\includegraphics[width=8.1cm, height=7.1cm]{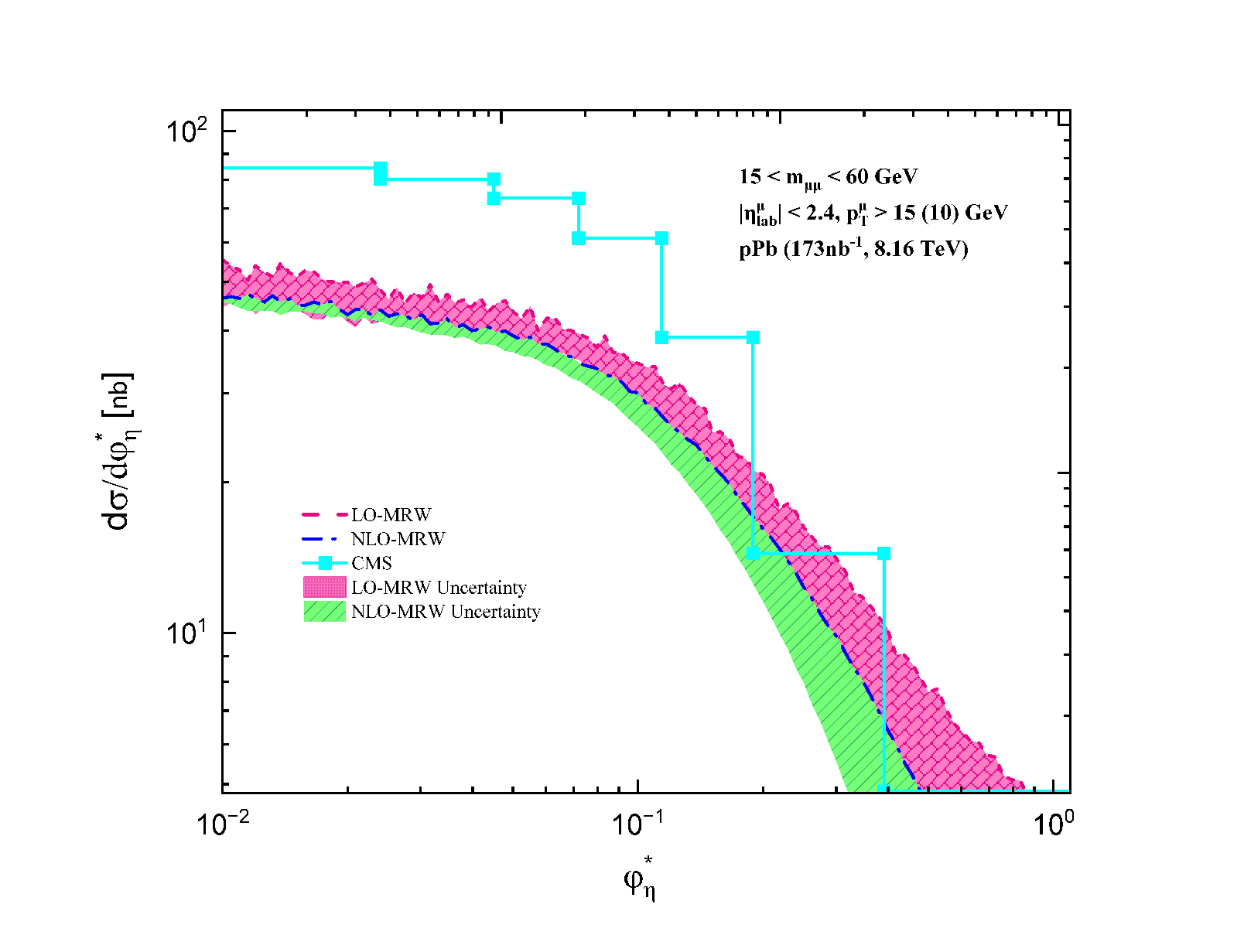}
	\includegraphics[width=8.4cm, height=7.1cm]{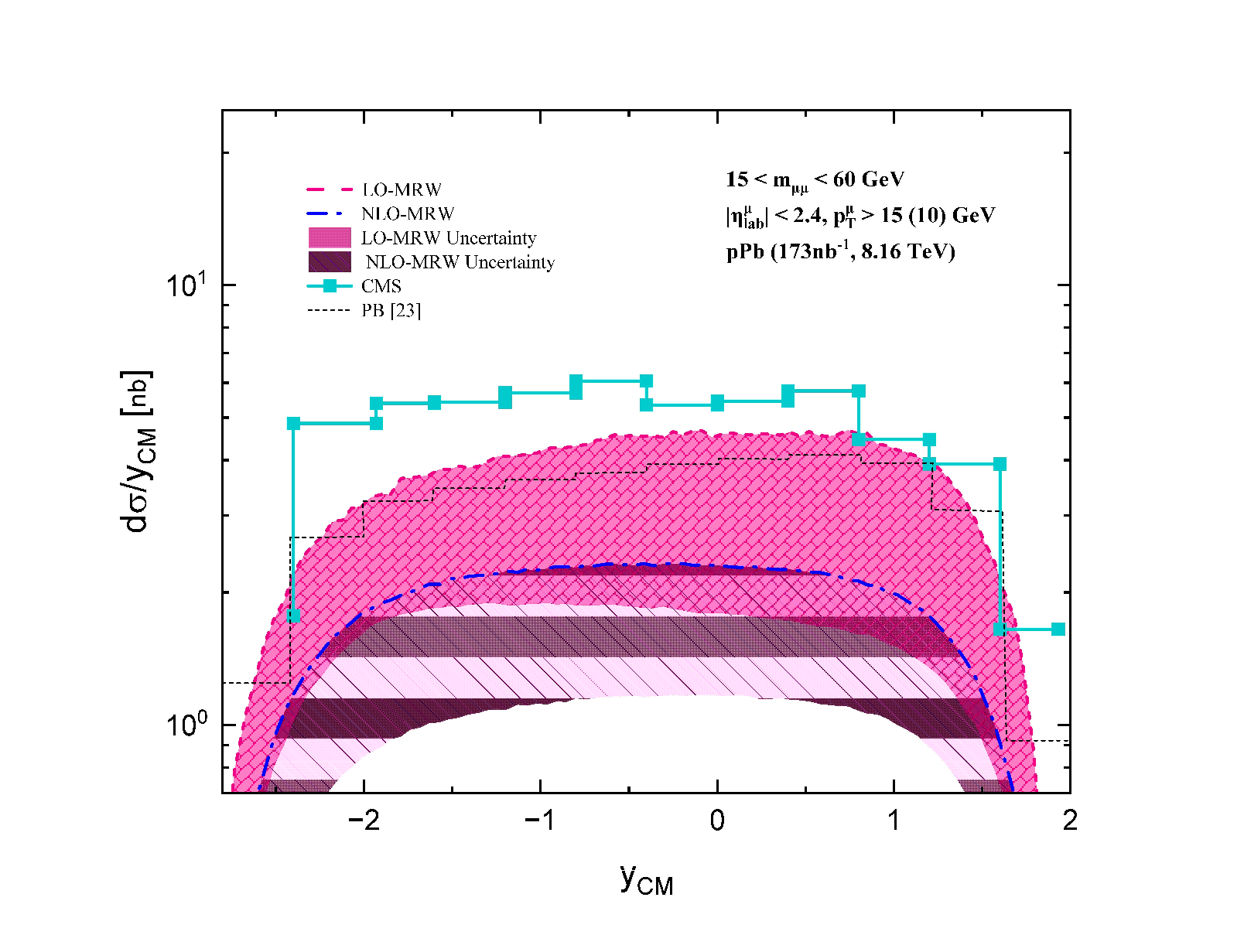}
	\caption
	{ Comparison of the differential cross sections with respect to  $P_t^{ll}$, (${\phi}^*_{\eta}$), and  $y_{CM}$ in the  $15 < m_{\mu\mu} < 60\;GeV$ interval  using LO-MRW, NLO-MRW and PB TMDs of the reference \cite{hanes}.}
	\label{fig:All-15-60}
\end{figure}
\begin{figure}
 	\centering
 	\includegraphics[width=8.1cm, height=7.1cm]{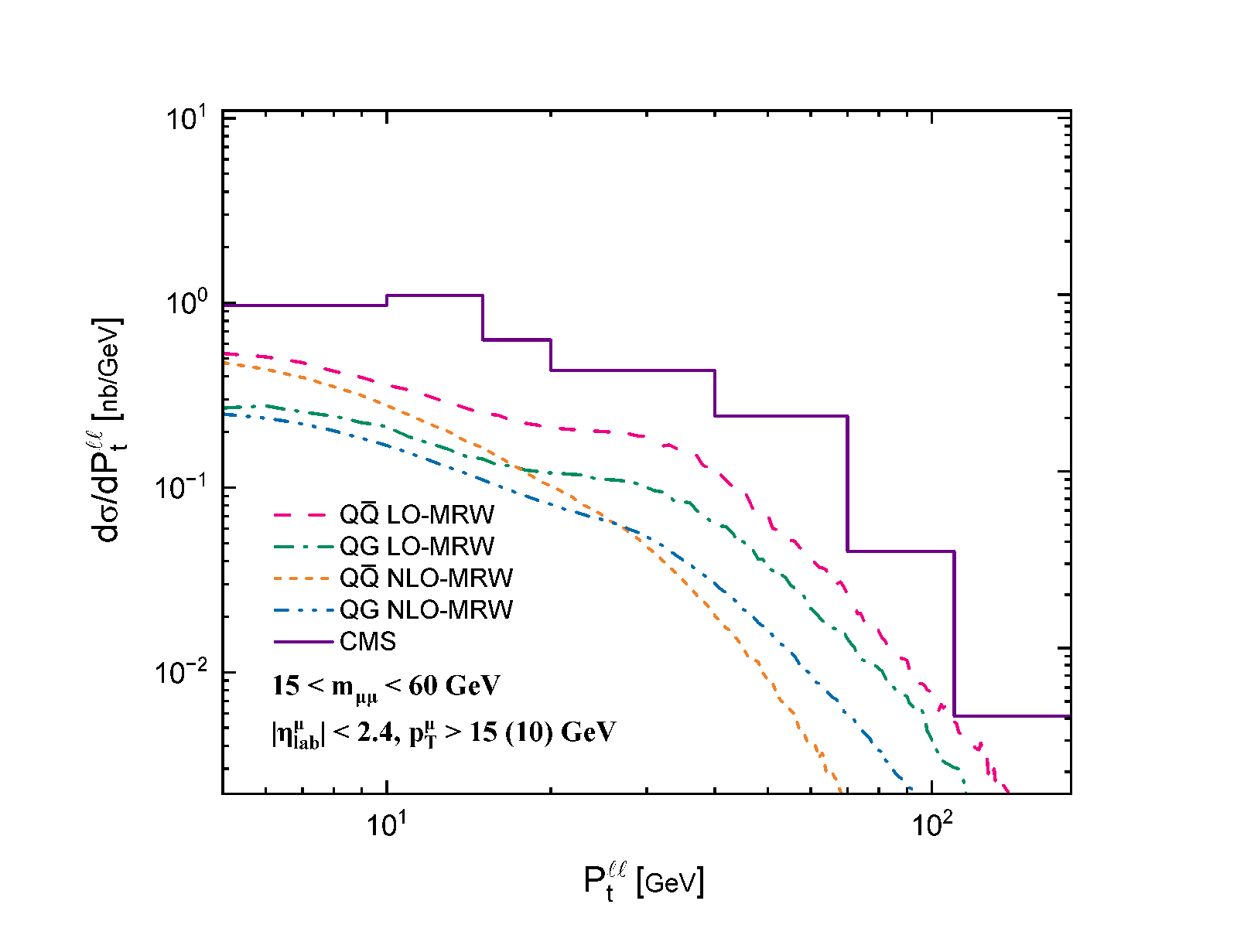}
 	\includegraphics[width=8.1cm, height=7.1cm]{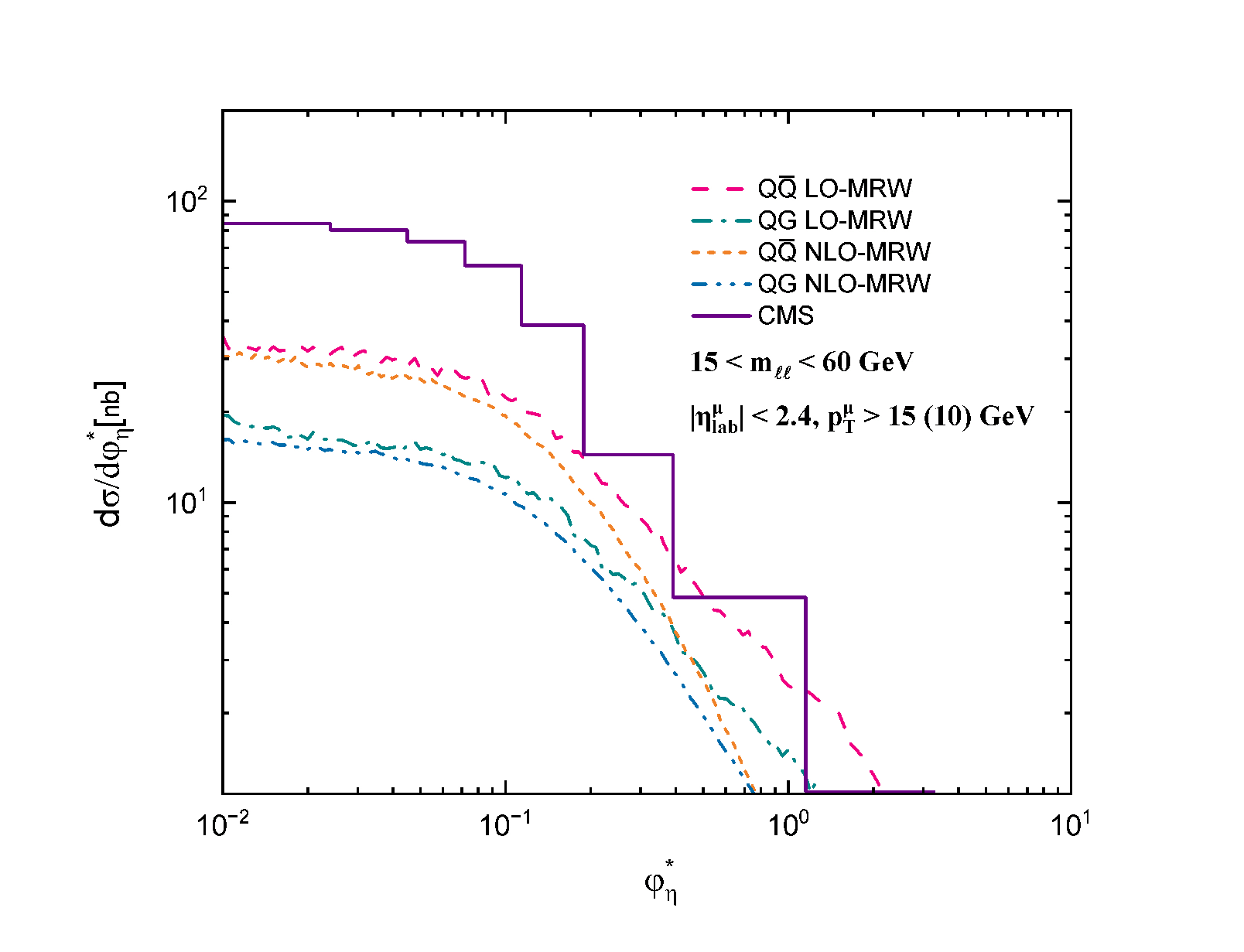}
 	\includegraphics[width=8.4cm, height=7.1cm]{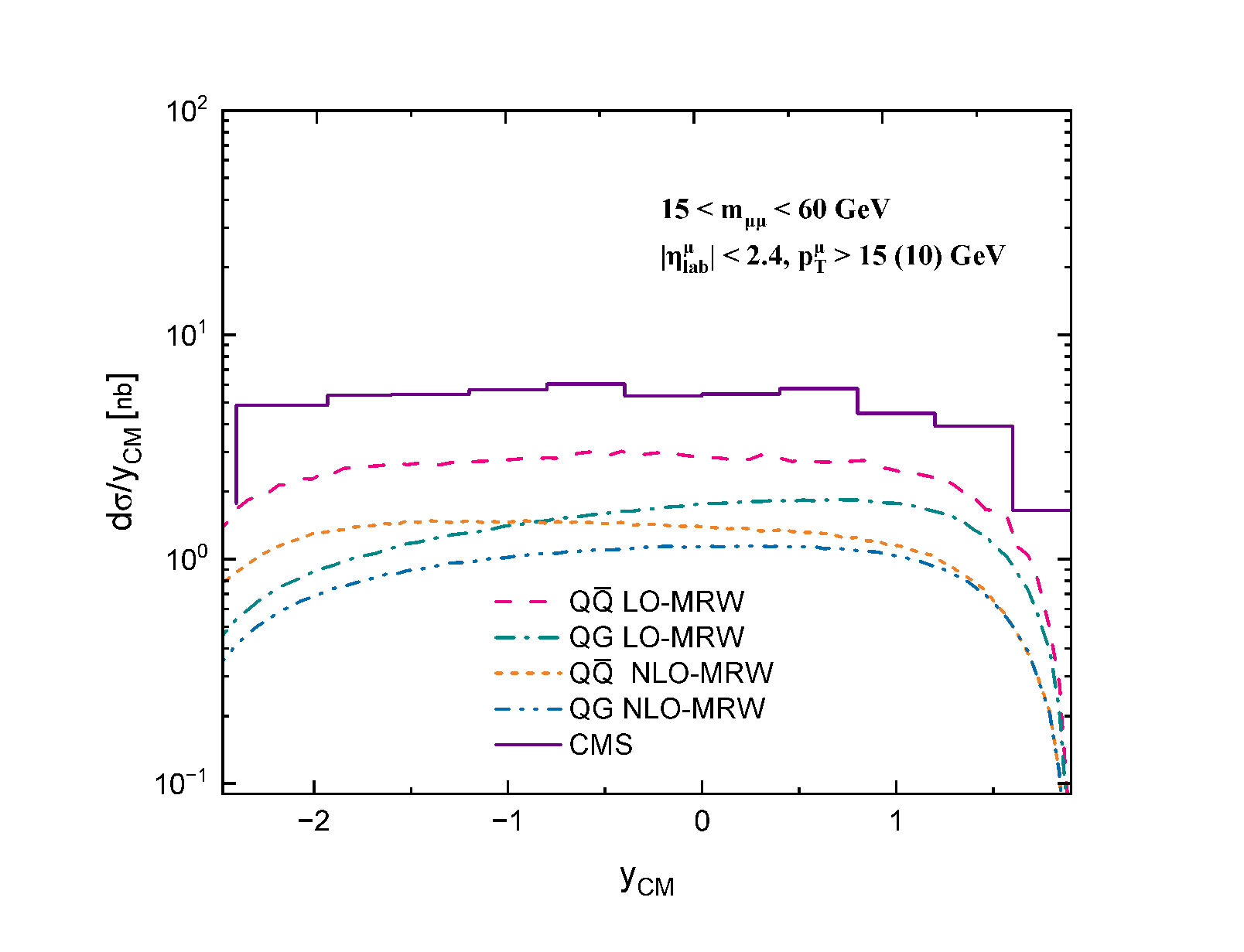}
 	\caption
 	{ Comparison of the differential cross sections of each subprocess with respect to  $P_t^{ll}$, (${\phi}^*_{\eta}$), and  $y_{CM}$ in the  $15 < m_{\mu\mu} < 60\;GeV$ interval  using LO-MRW, and NLO-MRW TMDs.}
 	\label{fig:Sub-15-60}
 \end{figure}
 In \cref{fig:Sub-60-120}, different subprocess contributions for both the LO-MRW and NLO-MRW TMD models are compared with respect to $P_{t}^{ll}$, ${\phi}^*_{\eta}$, and $y_{CM}$ kinematic variables in the $60 < m_{\mu\mu} < 120 \, \text{GeV}$ limit. The $q\overline{q} \rightarrow \ell^+ + \ell^-$ subprocess has a much larger contribution compared to $q+g \rightarrow \ell^+ + \ell^- + q$. In large ${\phi}^*_{\eta}$ and $P_t^{ll}$, the LO-MRW TMD model has larger contributions relative to the NLO-MRW model. Interestingly, NLO-MRW predictions for differential cross sections at large $p_T$ and ${\phi}^*_{\eta}$ for the $q+g \rightarrow \ell^+ + \ell^- + q$ subprocess become larger than the $q\overline{q} \rightarrow \ell^+ + \ell^-$ subprocess. Similar to the collinear factorization framework, gluonic contributions in the NLO-MRW framework become more dominant than quark distributions, at regions sensitive to large partonic transverse momenta, and small x limits, which is in contrast to the LO-MRW TMD model. This issue shows that in order to support data using the NLO-MRW TMD model at those regions, it is necessary to consider high order subprocesses similar to the collinear factorization framework.
 In \cref{fig:All-15-60}, we present a comparison of the differential cross section predictions using LO-MRW, NLO-MRW, and PB TMD models with respect to $P_t^{ll}$, ${\phi}^*_{\eta}$, and $y_{CM}$ kinematic variables in the $15 < m_{\mu\mu} < 60 \, \text{GeV}$ region. In the small dilepton mass region, higher order subprocesses are critical, but we only consider two lower order subprocesses, so we do not expect great results. It can be seen from these figures that the LO-MRW formalism prediction seems closer to the experimental data, while the NLO-MRW approach significantly undershoots it. Even the PB approach does not show good predictive power and undershoots experimental data, especially at low dilepton transverse momenta. Similar results are evident for differential cross sections with respect to ${\phi}^*_{\eta}$ and $y_{CM}$ where all TMD models predictions, undershoot experimental data. It should be mentioned that similar to what was observed and concluded for the NLO-MRW TMD models, here the predictions using this model becomes much more worse. Due to the fact that in this dilepton mass limit we are working the regions corresponding to high dilipeton transverse momenta regions, and only taking into account higher order subprocesses can improve the predictions using the NLO-MRW TMD model.
                             
  Finally, in \cref{fig:Sub-15-60}, where subprocesses of LO-MRW and NLO-MRW TMDs models are compared with respect to $P_{t}^{ll}$, ${\phi}^*_{\eta}$, and $y_{CM}$ kinematic variables in the $15 < m_{\mu\mu} < 60 \, \text{GeV}$ limit, a similar behavior for the TMD models predictions like what we discussed for the \cref{fig:Sub-60-120} are observed.

\section {Conclusion}
In this study, we have explored the production of Z bosons in proton-lead collisions using various transverse momentum dependent parton distribution functions (TMDs), specifically the LO-MRW, NLO-MRW, and PB approaches. Our analysis was conducted at a center of mass energy of $\sqrt{s} = 8.16$ TeV, and the results were compared against experimental data from the CMS collaboration.

Our findings indicate that both LO-MRW and NLO-MRW TMD models generally provide a good agreement with the experimental data, particularly in the $60 < m_{\mu\mu} < 120 \, \text{GeV}$ region. However, the LO-MRW model tends to have large predictions compared to  the data at high transverse momentum and ${\phi}^{\eta}$, while the NLO-MRW model, constrained by a strong ordering cutoff, shows better alignment in these regions. 

In the lower mass region of $15 < m_{\mu\mu} < 60 \, \text{GeV}$, the predictive power of the models is challenged, particularly for the NLO-MRW approach, which significantly undershoots the experimental data. This discrepancy highlights the importance of including higher order subprocesses.

Overall, our study underscores the effectiveness of the MRW approach in describing Z boson production in proton-lead collisions, while also highlighting the need for further refinement in TMD modeling, particularly in the low mass region.

\section*{ACKNOWLEDGEMENTS}
We would like to express our gratitude to the Institute for Research in Fundamental Sciences (IPM) for their financial support and services. We also extend our sincere appreciation to R. Kord Valeshabadi for his valuable comments.	
	%
\end{document}